\documentclass[aps,preprint,superscriptaddress]{revtex4-1}
\usepackage{color,bm,epsfig,graphics,amssymb,amsmath,subeqnarray,setspace,graphicx,amsthm,epstopdf,subfigure,color}
\usepackage{color,bm,epsfig,graphics,amssymb,amsmath,subeqnarray,setspace,graphicx,amsthm,epstopdf,subfigure,color,tipa,subfigure}
\usepackage[retainorgcmds]{IEEEtrantools}

\begin{document}

\title{Self-propulsion in viscoelastic fluids: pushers vs.~pullers }

\author{Lailai Zhu}
\email{lailai@mech.kth.se}
\affiliation{Linn\'e Flow Centre, KTH Mechanics, S-100 44 Stockholm, Sweden}

\author{Eric Lauga}
\affiliation{Department of Mechanical and Aerospace Engineering, University of California San Diego, 9500 Gilman Drive, La Jolla CA 92093-0411, USA.}

\author{Luca Brandt}
\affiliation{Linn\'e Flow Centre, KTH Mechanics, S-100 44 Stockholm, Sweden}

\date{\today}
  
\begin{abstract}
 
We use numerical simulations to address  locomotion at zero Reynolds number in viscoelastic (Giesekus) fluids. The
swimmers are assumed to be spherical, to self-propel using tangential surface deformation, and the computations are
implemented using a finite element method. The emphasis of the study is on the change of the swimming kinematics,
energetics, and flow disturbance from Newtonian to viscoelastic, and on the distinction between pusher and puller
swimmers.  In all cases,
the viscoelastic swimming speed is below the Newtonian one, with a minimum obtained for intermediate values of the
Weissenberg number, $We$. An analysis of the flow  field places the origin of this swimming degradation in non-Newtonian
elongational stresses. The power required for swimming is also  systematically below the Newtonian power, and always a
decreasing function of $We$. A detail energetic balance of the swimming problem points at the polymeric part of the
stress as the primary  $We$-decreasing energetic contribution, while the  contributions of the work done by the swimmer
from the solvent  remain essentially $We$-independent. In addition, we observe negative values of the polymeric power
density in some flow regions, indicating positive elastic work by the polymers on the fluid. The hydrodynamic
efficiency, defined as the ratio of the useful to total rate of work, is always above the Newtonian case, with a maximum
relative  value obtained at intermediate Weissenberg numbers. Finally, the presence of polymeric stresses leads to an
increase of the rate of decay of the flow velocity in the fluid, and a decrease of the magnitude of the stresslet
governing the magnitude of the effective bulk  stress in the fluid.

\end{abstract}
\maketitle

\section{Introduction}

The physical mechanisms of microorganism locomotion accompany a variety of natural phenomena, including human
spermatozoa approaching the ovum in the mammalian female reproductive tract \cite{fauci06}, algal blooms moving to
nutrient rich environment in maritime regions \cite{pedley92,hill05}, biofilm formation \cite{costerton87},  and
paramecia cells escaping from  their predators \cite{baroud}. Much work has been done to shed light on the different
physical mechanisms at play in  the locomotion of microorganisms, including
classical kinematics studies~\cite{brennen77,PRLswim, LaugaSwim, Sphere-envelope}, nutrients
uptake~\cite{nutrientPedley,feedingMichelin}, collective
behavior~\cite{cisneros07,hernandez-ortiz05,PhysRevLett.100.088103,Moorespermcoll,saintillan_POF} and hydrodynamic
interactions~\cite{swimInter-Pedley, IshikawaUnsteady,Ishikawa_TwoProlate}. 

Most past work has however been limited to the study of locomotion in Newtonian
fluids, whereas in many biological instances microorganisms swim in complex
polymeric fluids.  Several groups have recently started to address the effect of
viscoelasticity on the locomotory features and hydrodynamic performance of
swimming microorganisms. Lauga~\cite{lauga07} investigated the
kinematics and energetics of Taylor's waving sheet in various model nonlinear
polymeric fluids and found that, for a given waving stroke, viscoelasticity
impeded the locomotion of the swimming sheet. Similarly,  Fu et
al.~\cite{Henry-Charles} derived analytically that the swimming speed of an
infinite filament with a prescribed waveform will decrease in Oldroyd-B fluid
compared to the Newtonian case. Teran et al.~\cite{PhysRevLett.104.038101}
numerically studied the influence of viscoelasticity on the dynamics of a
finite-length Taylor's waving sheet and showed that the swimming speed and
efficiency can actually increase within a certain range of polymer relaxation
time and sheet waveforms. Recently, Shen et al.~\cite{PRL_Shen_Visco}
experimentally demonstrated that fluid elasticity hindered the self-propulsion
of the nematode \textit{Caenorhabditis elegans}, primarily due to  polymeric
stretching near hyperbolic points in the viscoelastic flow.

In our previous work~\cite{lailai-pre}, we carried out three-dimensional
axisymmetric simulations to study the effect of viscoelasticity on the
locomotion speed and energy consumption of a spherical squirmer swimming by
tangential surface deformations, as a model of ciliated protozoa. A non-linear
viscoelastic model (Giesekus) was used to describe the polymer dynamics. We
showed the dependency of swimming speed, power consumption, and swimming
efficiency on the fluid elasticity. This  work was however restricted to the most
energetically efficient swimming gait, one associated with no vortices generated
around the cell in the Newtonian case (a potential flow squirmer in the Newtonian limit). 

In the current paper we numerically investigate  the hydrodynamic performance of a
squirmer utilizing other swimming gaits in a viscoelastic fluid. The gaits
correspond to pusher and puller swimmers, a natural distinction for biological
organisms. We assume the gait to be steady and  thus ignore, as a first model,  the oscillatory
flow field around a swimming cell. As recently shown by Guasto et al.~\cite{JeffPrlOsci} and Drescher et al.~\cite{DrescherPrl}, the flow field generated by some microorganisms such as \textit{Chlamydomonas reinhardtii} cannot be accurately described by a steady analysis, and in fact, the flow generated by many swimmers is usually time-periodic, with large fluctuations around the mean. However, the simple steady model we utilize here has been used to understand several fundamental processes related to the physics of swimming microorganisms, such as nutrient uptake~\cite{nutrientPedleysteady}, locomotion in stratified 
fluid~\cite{swimPycnoclines}, biomixing~\cite{biostirring} and the collective behavior of microorganisms~\cite{PhysRevLett.100.088103,Art_squirmer}. Note that in the case of viscoelastic fluids like those studied here, the presence of an additional time scale (stroke frequency) leads to the definition of a Deborah number, different from the Weissenberg number defined below based on shear rates, and possibly leading to a rich dynamical behavior.

The paper is organized as follows. We first investigate in detail how fluid elasticity affects the
swimming speed of the spherical model cells applying different locomotive gaits and the resulting
polymeric dynamics in the flow field. We then analytically derive a
decomposition of the power associated to cell locomotion by surface deformation
in the viscoelastic fluid, extending the analysis presented for the Newtonian
fluid in Ref.~\cite{PRLswim}. We further compute the change in the swimming
speed of the squirmer in the viscoelastic fluid under constant-power conditions, as well as the hydrodynamic efficiency. 
Finally, in order to quantify the influence of the squirming motion on nearby
swimmers as well as on the bulk stress of the surrounding fluid, we analyze the
velocity decay around the model swimming cells and calculate their effective stresslets.

\section{Mathematical Model}\label{mathmodel} 

\subsection{Squirmer Model}
 In this so-called envelope
approach, first proposed by Blake \cite{Blake1971a}, one assumes the squirmer imposes a non-zero tangential velocity on
its surface, $\mathbf{u}_S$, in the co-moving frame. Here we adopt the concise formulation introduced in
Ref.~\cite{PhysRevLett.100.088103}
\begin{equation}
 \mathbf{u}_{\small{S}} ({\bf r})=\sum_{n\geq1} \frac{2}{n(n+1)}B_n
  P_{n}^{'}\left(\frac{\bf{e} \cdot \mathbf{r}}{r}\right)
  \left( \frac{\mathbf{e} \cdot \mathbf{r}}{r} \frac{\mathbf{r}}{r}-\mathbf{e} \right),
\end{equation}
where $\mathbf{e}$ is the orientation vector of the squirmer, $B_n$ is the $n$th mode of the surface squirming velocity \cite{Blake1971a}, $P_{n}$ is the $n$th Legendre polynomial, $\mathbf{r}$ is the position vector, and $r=|\mathbf{r}|$. 
In a Newtonian fluid, the swimming speed of the squirmer is $U_{\text{New}}=2B_{1}/3$ \cite{Blake1971a} and thus only
dictated by the first mode. As in many previous
studies~\cite{swimInter-Pedley,nutrientPedley,Ishikawa_TwoProlate,PhysRevLett.100.088103}, we assume $B_{n}=0$ for
$n\geq 3$.
The tangential velocity on the sphere in the co-moving frame is therefore expressed as $u_{\theta}(\theta)=B_{1}\sin\theta+(B_{2}/2 )\sin2\theta$, where $\theta=\arccos (\mathbf{e} \cdot \mathbf{r}/r )$.  We introduce an additional parameter $\alpha$, representing the
ratio of the second to the first squirming mode, thus $\alpha =B_{2}/B_{1}$.
When $\alpha$ is positive, the swimmer get impetus from its front part, and is termed  a puller (swimmers in this category include the alga genus {\it Chlamydomonas}). As a difference, when $\alpha$ is negative, the thrust comes from the rear part of the body and the swimmer is a pusher (swimmers in this category include all peritrichous bacteria such as {\it Escherichia coli}). If not otherwise stated, here we will present results obtained with $\alpha=-5$ (simply denoted as pusher),  $\alpha=5$ (puller) and $\alpha=0$ (neutral squirmer inducing a potential velocity field in the Newtonian case).

\subsection{Polymeric Dynamics}
For steady incompressible low-Reynolds number flow in a viscoelastic fluid,  the momentum and
continuity equation are written as
\begin{eqnarray} \label{eq:mom}
 -\nabla p + \nabla \cdot \boldsymbol{\tau}&=&0,\\ 
 \nabla \cdot \mathbf{u}&=&0.\label{eq:con}
\end{eqnarray}
The velocity is scaled by $B_{1}$ -- related by the expressions above to the magnitude of the velocity of the boundary and to the swimming speed -- while lengths are scaled by the diameter of the spherical swimmer $D$, time is scaled by $D/B_{1}$, and pressure and stresses are scaled by $\mu B_{1} / D$, where $\mu$ is the coefficient of viscosity. 

Following classical modeling approaches \cite{birdvol1,birdvol2,bird76}, 
the deviatoric stress $\boldsymbol{\tau}$ is split into two components, the viscous solvent stress
($\boldsymbol{\tau}^{s}$) and the polymeric stress ($\boldsymbol{\tau}^{p}$), so
$\boldsymbol{\tau}=\boldsymbol{\tau}^{s}+\boldsymbol{\tau}^{p}$.
The stress $\boldsymbol{\tau}^{s}$ is governed by Newton's law of viscosity, thus is formulated as
\begin{equation}
 \boldsymbol{\tau}^{s}=\beta \left(\nabla \mathbf{u} + \nabla \mathbf{u}^{T}\right),
\end{equation}
where $\beta < 1$ represents the ratio of the solvent viscosity, $\mu_{s}$, to the total zero shear rate viscosity, 
$\mu$, of the solution.
To close the model, a transport equation for the polymeric stress $\boldsymbol{\tau}^{p}$ is required. Here we adopt the nonlinear Giesekus model \cite{giesekus}, which, in addition to displaying shear-thinning material properties (viscosity and normal stress differences),  provides two important features, namely saturation of polymer elongation, and a non-negative entropy production during the time evolution of the polymers (see details in Ref.~\cite{Beris-Edwards,Dupret-Marchal,Souvaliotis-Beris}). Violation of these two properties may cause non-physical flow behavior as well as numerical difficulties. The nondimensional constitutive equation for this model can be written as
\begin{equation}\label{eq:poly}
 {\boldsymbol{\tau}^p}+We \, \overset{\triangledown}{\boldsymbol{\tau}^p} + \frac{We \, \, \alpha_m}{1-\beta}(\boldsymbol{\tau}^p \cdot\boldsymbol{\tau}^p)=\ (1-\beta\ )(\nabla \mathbf{u} + \nabla \mathbf{u}^{T}),
\end{equation}
where $ \overset{\triangledown}{\mathbf{A}}$ denotes the upper-convected derivative,  defined for a tensor $\mathbf{A}$ by
\begin{equation}
 \overset{\triangledown}{\mathbf{A}}=\frac{\partial \mathbf{A}}{\partial t}+\mathbf{u} \cdot \nabla \mathbf{A}-\nabla \mathbf{u} ^T \cdot \mathbf{A}-\mathbf{A} \cdot \nabla \mathbf{u}.
\end{equation}
In the constitutive equation for the polymeric stress, Eq.~\eqref{eq:poly}, $We$ denotes the Weissenberg number, defined as $We={\lambda B_{1}}/{D}$, where $\lambda$ is the polymer relaxation time. The so-called mobility factor $\alpha_m$ is introduced in the nonlinear stress term to represent an anisotropic hydrodynamic drag on the polymer molecules \cite{birdvol1} and limits the extensional viscosity of the fluid. From thermodynamics considerations, the mobility factor $\alpha_m$ must be in the 0 to 1/2 range  \cite{birdvol1,Larson}, and in this paper we assume $\alpha_m=0.2$.

\section{Numerical method}

A numerically stable and accurate finite element model is built, based on the formulation denoted Discrete Elastic-Viscous Split Stress (DEVSS-G)~\cite{devssFortin,Devss-G98}.
The governing equations are rewritten as
\begin{equation}\label{devssg}
 \nabla \cdot \mu_{a}(\nabla \mathbf{u} + \nabla \mathbf{u}^{T})-\nabla p + \nabla \cdot
\boldsymbol{\tau}^{p}-\nabla \cdot (\mu_{a} - \beta)(\mathbf{G} + \mathbf{G}^{T})=0,
\end{equation}
where an additional tensor $\mathbf{G} \equiv \nabla \mathbf{u}$ is introduced as an independent interpolation of the
velocity gradient tensor $\nabla \mathbf{u}$ and an additional elliptic term $\nabla \cdot \mu_{a}(\nabla \mathbf{u} +
\nabla \mathbf{u}^{T}) - \nabla \cdot \mu_{a}(\mathbf{G} + \mathbf{G}^{T})$ is added into the
momentum equation \cite{MIT99_DAVSSDG}.
In our computations, $\mu_{a}$ is chosen to be $1$ as in Liu  et al.~\cite{Devss-G98}, and the velocity gradient
term in equation for the polymeric stress, Eq.~(\ref{eq:poly}), is approximated by $\mathbf{G}$.

A Galerkin method is used to discretize the momentum equations, continuity equation, and the equation for the additional
unknown $\mathbf{G}$. We use quadratic elements for $\mathbf{u}$ and linear elements for both $p$ and $\mathbf{G}$. To
improve the numerical stability,
the streamline-upwind/Petrov-Galerkin(SUPG)\cite{supgCrochet} method is used to discretize the constitutive equation,
Eq.~\eqref{eq:poly}. Finally, the weak form for the constitutive equations is written as
\begin{IEEEeqnarray}{rl}\label{weakpoly}
  \Big\{ \mathbf{S}+\frac{h}{U_C}\mathbf{u} \cdot \nabla \mathbf{S},\quad \boldsymbol{\tau}^{p} & + We (\mathbf{u} \cdot \nabla \boldsymbol{\tau}^p - \mathbf{G}^{T} \cdot \boldsymbol{\tau}^{p} - \boldsymbol{\tau}^{p} \cdot \mathbf{G}  + \frac{\alpha_{m}}{1-\beta}(\boldsymbol{\tau}^{p} \cdot \boldsymbol{\tau}^{p}))
  \nonumber \\
&-\ (1-\beta\ )(\nabla \mathbf{u}+\nabla \mathbf{u}^T )\Big\}=0,
\end{IEEEeqnarray}
where $\mathbf{S}$ denotes the test function for $\boldsymbol{\tau}^{p}$, $h$ is the characteristic length-scale of the element and $U_C$ is the magnitude of the local characteristic velocity. In our case, we choose the norm of $\mathbf{u}$ as the value of $U_C$.
The finite-element framework for our implementation is provided by the commercial software COMSOL.

We perform
three-dimensional axisymmetric simulations on a two-dimensional mesh.
Triangle elements are used for all the simulations and the number of elements
are dependent on the swimming gait.
We generate
sufficiently refined mesh in regions with high magnitude of polymeric stress  to
overcome numerical instabilities~\cite{viscoeReview,computaRheologyCFD}.
In all cases, high resolution near the cell body is required in order to resolve the thin stress boundary layers.
Fine meshes are also necessary to capture the high polymer elongation rates. Typical mesh size at the cell boundary is
$5E10^{-4}$ of the
reference (swimmer) length. The total  number of elements used for the results reported
here ranges from $150,000$ to $250,000$.  

The numerical implementation is  first validated against the theoretical results for the swimming  speed and power
of squirmers of different gaits in Newtonian fluids~\cite{Blake1971a}.
For viscoelastic fluids, we validate the present
numerical approach against the simulations in Ref.~\cite{Lunsmann1993} of a sphere
sedimenting in a tube filled with an Oldroyd-B fluid (the mobility factor $\alpha_m$ equals to
zero) and against the results in Ref.~\cite{harlen_giesekus_wake} where the authors
numerically study the negative wake of a sphere sedimenting in a tube filled with a
Giesekus fluid.

\section{Swimming speed  in viscoelastic fluids}\label{subsec:speed}

\begin{figure}[t]
 \begin{center}
  \includegraphics[width=0.98 \textwidth]{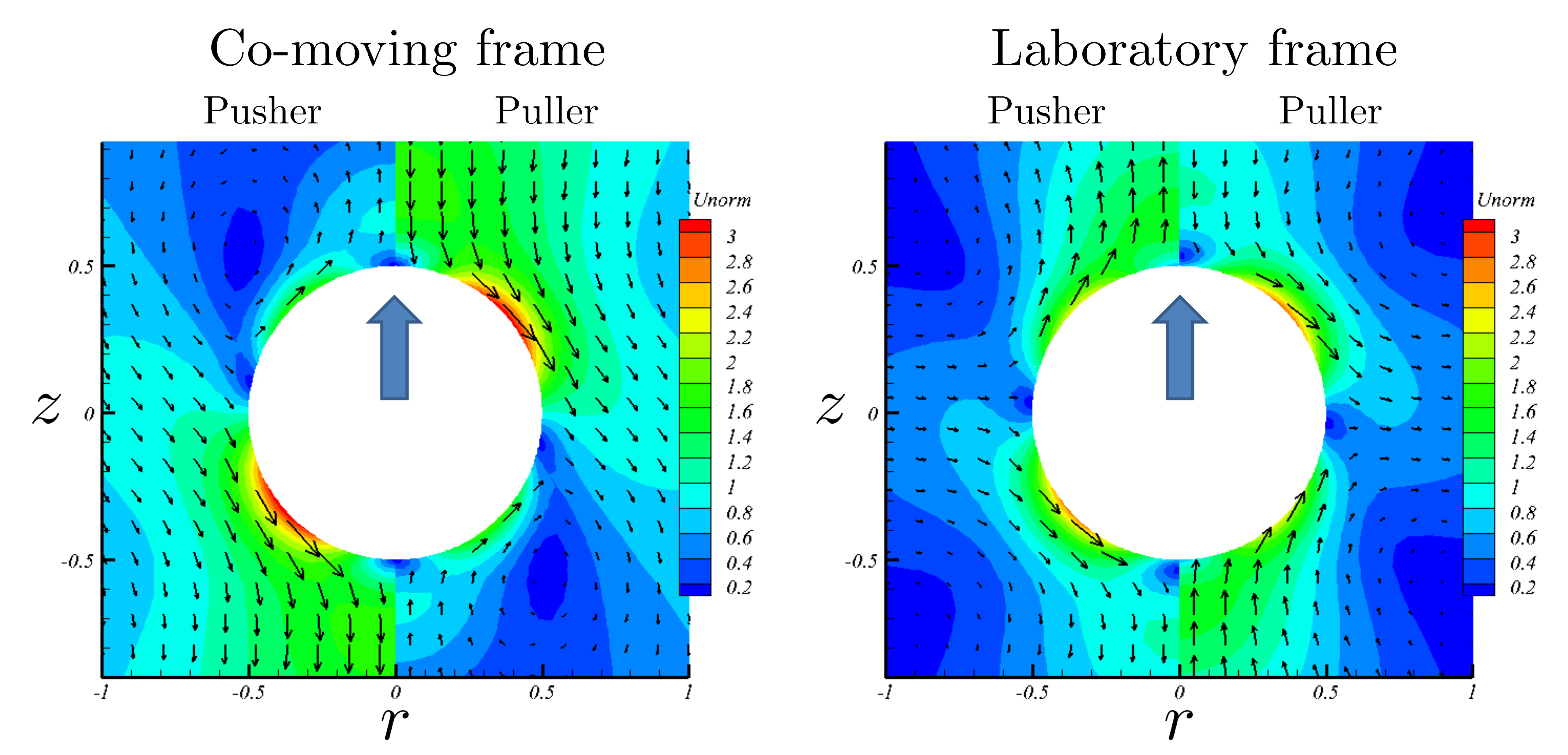}
 \end{center}
\caption{(Color online) Vector plot of flow fields generated by  pusher and  puller squirmers in a Newtonian fluid.
Left: co-moving frame. Right: laboratory frame. On each panel,  the data for the pusher are displayed on the left, and
those for the puller on the right. The large arrow (blue online) indicates the swimming direction and
the background color scheme indicates the velocity magnitudes.}
\label{fig:flow_pushpull}
\end{figure}

To illustrate the difference between the swimming modes considered, the flow fields generated by a pusher and a puller in a Newtonian fluid are shown in Fig.~\ref{fig:flow_pushpull}, where data for the pusher are displayed on the left of each panel, and those for the puller on the right. Vectors display the in-plane velocity magnitude and direction while the background color indicates velocity magnitudes. We see an axisymmetric vortex  ring generated in the front of the pusher and behind  the puller; as a difference the neutral swimmer generates a symmetric velocity field with no vorticity~\cite{lailai-pre}.

\begin{figure}[t]
 \begin{center}
  {\includegraphics[scale=0.28]{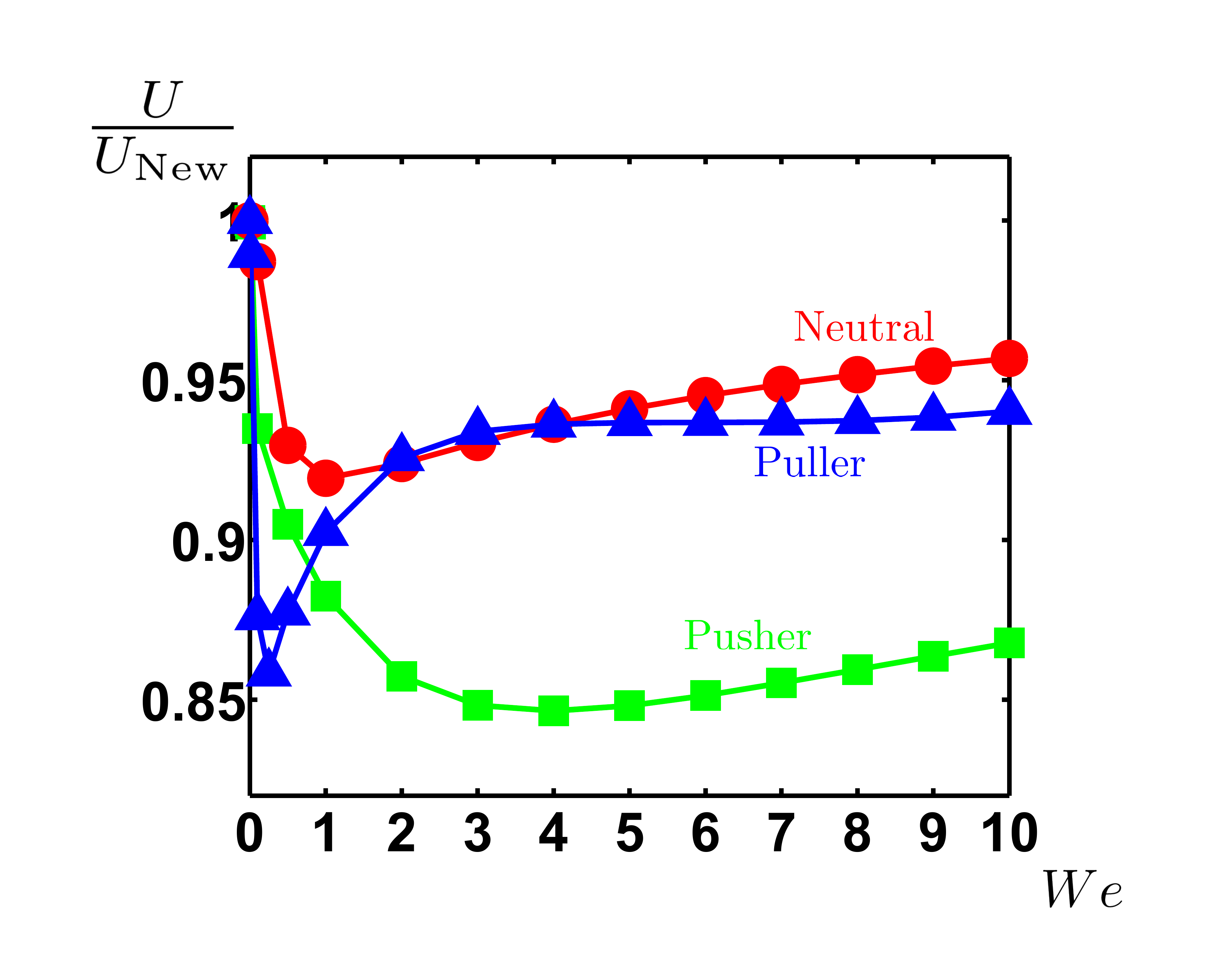}}
 \end{center}
\caption{(Color online) Nondimensional swimming speed, $U$, of the swimmer as a function of the Weissenberg number,
$We$, for pusher $\alpha=-5$ (squares, green online), puller  $\alpha=5$ (triangles, blue online) and neutral
squirmer $\alpha=0$ (circles, red online). The swimming speed is scaled by its corresponding value in the Newtonian
fluid, $U_{\rm New}$.}
\label{fig:AR1_spd}
\end{figure}

To address the dynamics in a viscoelastic fluid, we first examine the dependency of the squirmer swimming speed, $U$, on the Weissenberg  number, $We$. 
The swimming speed is computed as in Ref.~\cite{lailai-pre}. Simulations are performed in the co-moving frame with different values of the inflow velocity, and the total force on the body is computed. The swimming speed is then estimated by interpolating to the inflow velocity yielding zero net force on the swimmer. A new simulation is then performed to ensure that the value of the predicted swimming speed used as free-stream velocity indeed gives values of the total force below a given tolerance. The numerical results are shown in Fig.~\ref{fig:AR1_spd}, for three different swimming gaits, where the swimming speed is scaled with the speed of the Newtonian squirmer, $U_{\rm New}=2 B_1/3$.

Considering the data in Fig.~\ref{fig:AR1_spd}, we first observe that
viscoelasticity clearly distinguishes between the three swimming gaits in terms
of $U$, while they produce the same swimming speed in the Newtonian fluid.  The
neutral  gait, $\alpha=0$,  generally gives the highest value of $U$ for most
$We$ numbers while  pushers generally have the lowest velocity. Second, for all
the gaits and $We$ numbers considered here, the swimming speed of swimmers in
the viscoelastic fluid is lower than that of swimmers in the Newtonian fluid.
Finally, for all three gaits, $U$ systematically displays a minimum in the range
of $We$ considered.  The swimming speed initially decreases as $We$ increases
from $0$, while it eventually always increase for the largest values of $We$.
The value of the Weissenberg  number corresponding to the minimum swimming speed
varies with the gait: it increases from $0.5$ to $4$ when moving from puller to
neutral to pusher.

\begin{figure}
 \begin{center}
  \includegraphics[width=0.6\textwidth]{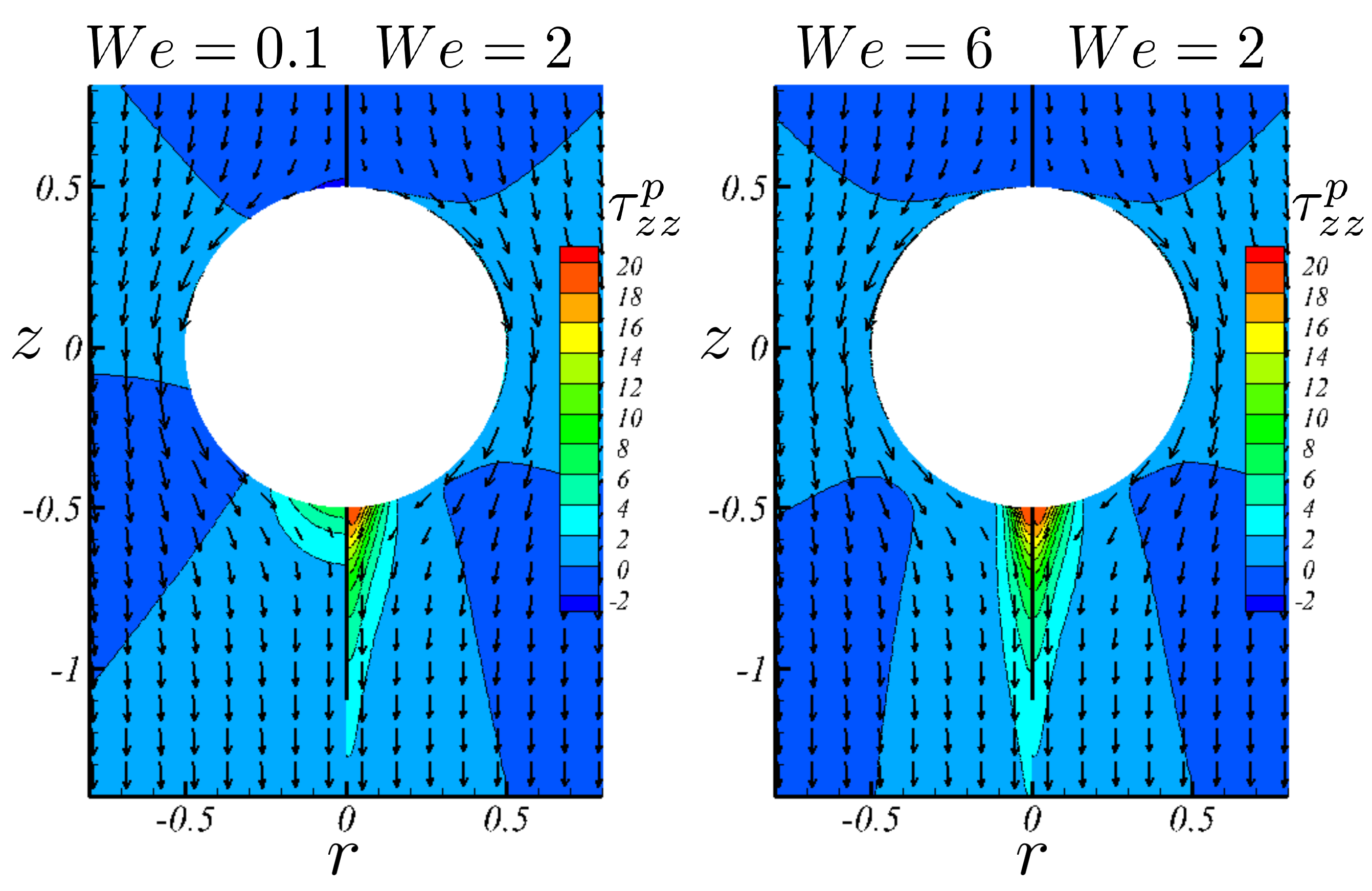}
 \end{center}
\caption{(Color online) Distribution of the $\tau_{zz}$ component of the
 polymeric stress for the neutral squirmer and three values of the Weissenberg
number, $We=0.1, 2$, and 6. (Left: $We=0.1$ and 2. Right: $We=6$ and 2.)}
\label{fig:tzz_al0_we01_2_we6_2}
\end{figure}

Different swimming gaits generate different polymer dynamics around the swimmer
and this,
in turn, influences the swimming speed as shown above. To gain insight into the
numerical results, we start by showing in Fig.~\ref{fig:tzz_al0_we01_2_we6_2}
the distribution of  axial polymeric stress, $\tau_{zz}$, for the neutral
 swimmer ($\alpha=0$) in the cases $We=0.1$ and $We=2$. The value $We=0.1$ is
chosen since it is close to the Newtonian case ($U= 0.987$), while $U$ is close
to the minimum swimming speed when $We=2$. As seen in the figure, there exists a
significant difference in the $\tau_{zz}$ distribution behind the swimmer. In
the case $We=2$, the magnitude of $\tau_{zz}$ is much higher, indicating a
higher extent of polymer stretching in this region. 
As suggested in 
Fig.~\ref{fig:flow_pushpull}, an elongational flow~\cite{birdvol1} is generated
aft the cell body. In the viscoelastic fluid such flow would be responsible for polymer streching which in return
increases the local elongational viscosity  \cite{elongationalvis1,elongationalvis2}, yielding a strong elastic resistance against locomotion. We further see in Fig.~\ref{fig:tzz_al0_we01_2_we6_2} that,
upon increasing the polymer relaxation time to $We=6$, the region of largest
elongation becomes narrower and closer to the pole, leading to a reduced elastic
resistance and consistent with the results of Fig.~\ref{fig:AR1_spd}.

\begin{figure}[b]
 \begin{center}
  \includegraphics[width=0.6\textwidth]{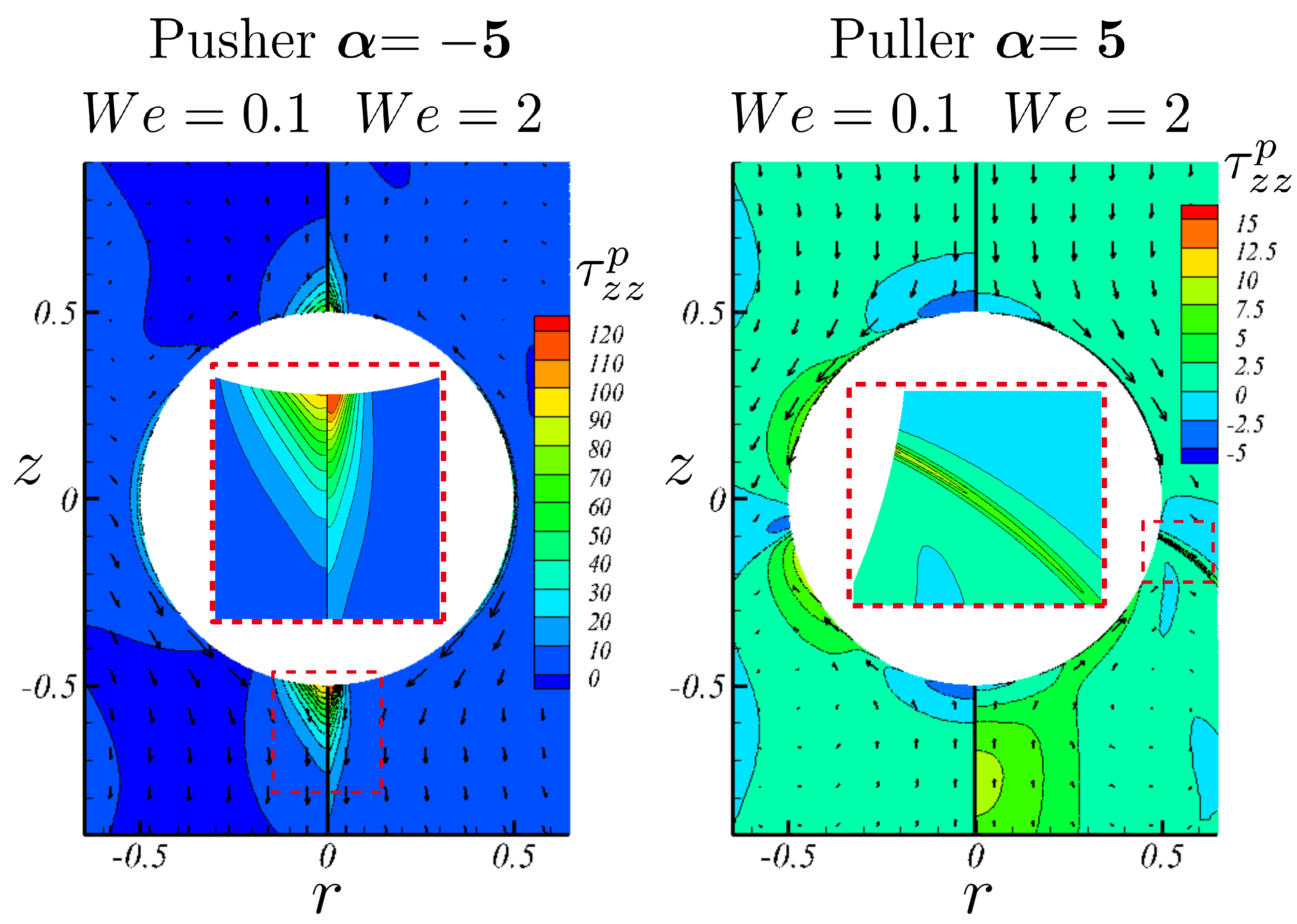}
 \end{center}
\caption{(Color online) Distribution of the $\tau_{zz}$
component of the polymeric stress for the pusher (left) and puller (right) swimming in the
viscoelastic fluid. For both gaits, $We=0.1$ and $We=2$ are chosen for
comparison. The inset plots show regions with high value of $\tau_{zz}$.
}
\label{fig:tzz_al-5_al5_we01_2}
\end{figure}

\begin{figure}[t]
 \begin{center}
  \includegraphics[width=0.55\textwidth]{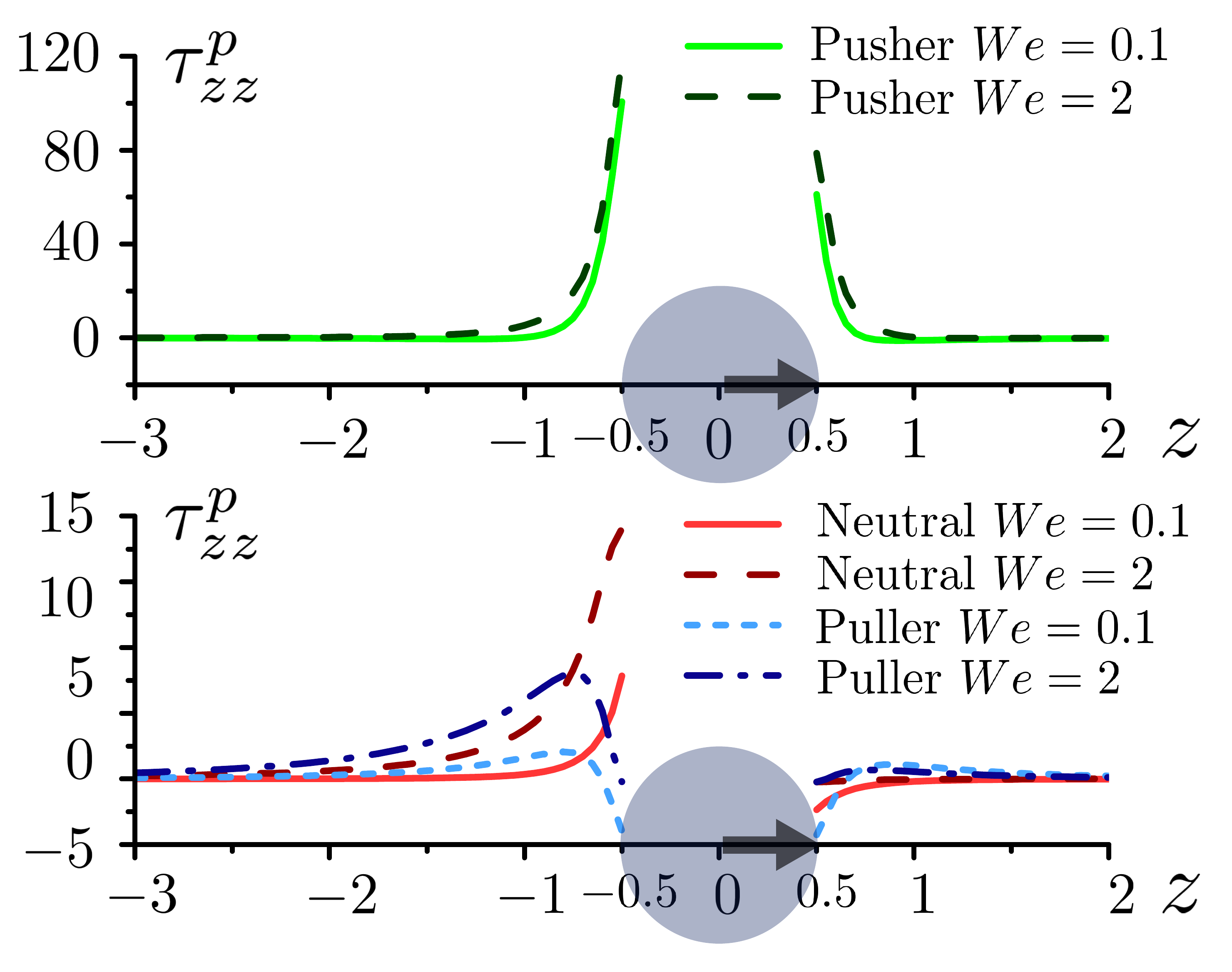}
 \end{center}
\caption{(Color online)  Distribution of the $\tau_{zz}$
component of the polymeric stress along the symmetry axis. Shaded circles
and arrows indicate the squirmer and its swimming direction. Data for the pusher are shown in the top 
whereas the neutral squirmer
and the puller are reported at the bottom. The solid and short-dashed lines correspond to
$We=0.1$, the long-dashed and dot-dashed lines to $We=2$.}
\label{fig:tzz_alongaxis_Allthreegaits}
\end{figure}

The polymer contribution for the pusher swimmer is displayed in
Fig.~\ref{fig:tzz_al-5_al5_we01_2} (left).
In this case, the kinematics of surface deformation {draws
fluid from the side and pushes it towards the two poles, resulting in strong
elongational flows at the poles (see Fig.~\ref{fig:flow_pushpull}). Such flow aligns the polymer chains near the poles in the swimming
direction, leading to high elongational viscosities in the front of and behind the
squirmer.} Comparing Fig.~\ref{fig:tzz_al0_we01_2_we6_2} and
Fig.~\ref{fig:tzz_al-5_al5_we01_2}, we see that the magnitude of $\tau_{zz}$ on
the back is much larger in the case of pushers. This is further 
further illustrated in Fig.~\ref{fig:tzz_alongaxis_Allthreegaits} where we display the value of the normal  polymeric stress along the symmetry axis.  
It is noteworthy that for the pusher swimmer, significant
$\tau_{zz}$ is evident also in the region ahead of the body. 
In this case, the stretched polymers might contribute  an elastic driving force on
the swimmer. However, the magnitude of $\tau_{zz}$ at the front is lower than
that behind, and thus the net effect is that of an additional drag. Note that for the neutral
squirmer, the region of polymer elongation becomes narrower at large $We$ and we observe a recovery of the swimming
speed.

\begin{figure}
 \begin{center}
  \includegraphics[width=0.55 \textwidth]{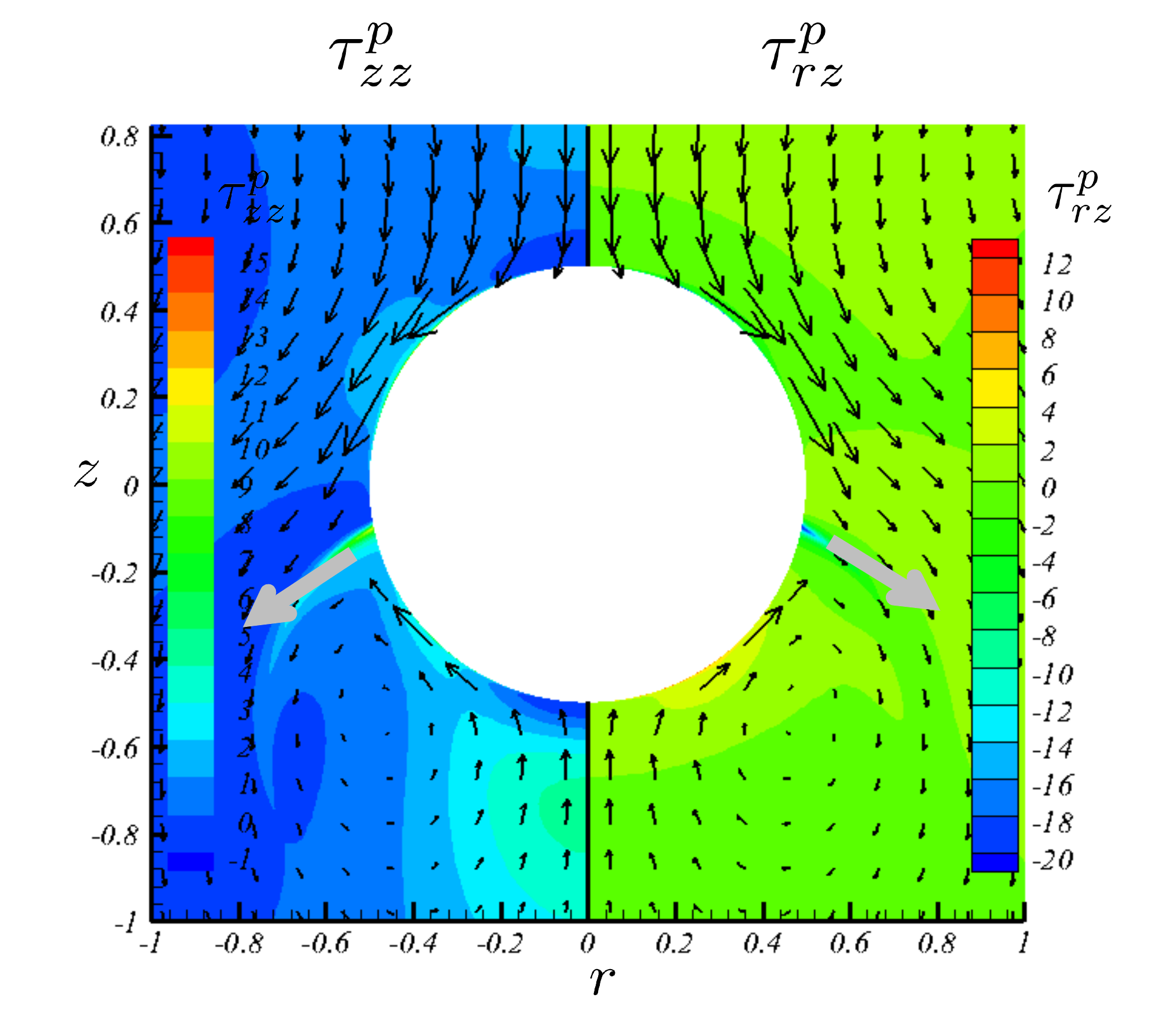}
 \end{center}
\caption{(Color online) Distribution of the $\tau_{zz}$ and $\tau_{rz}$
components of the polymeric stress for pullers in the viscoelasic fluid with
$We=0.5$. Both the $\tau_{zz}$ and $\tau_{rz}$ components lead to a net
polymeric (elastic)  force indicated by the gray arrows,  hindering  locomotion.
The location of this  elastic forces is right above the  vortex ring, implying
that they are generated by the elongational flow. }
\label{fig:tzz_trz_al5_we0.5}
\end{figure}

We then consider the puller swimmer, which is about 20\% faster than the  pusher (as seen in Fig.~\ref{fig:AR1_spd}).  The polymeric stresses in this case are shown in  Figs.~\ref{fig:tzz_al-5_al5_we01_2} (right), \ref{fig:tzz_alongaxis_Allthreegaits} and \ref{fig:tzz_trz_al5_we0.5}. The results in Fig.~\ref{fig:tzz_trz_al5_we0.5} are displayed for $We=0.5$, corresponding to the minimal swimming speed for the viscoelastic puller (and
thus the maximum influence of viscoelastic stresses). In the case of pullers,
the inward velocity imposed by the surface deformation on the front and back of
the body takes fluid away from  the back of the swimmer, and thus stretched fluid is removed from the region behind the body.  
As shown in Fig.~\ref{fig:tzz_alongaxis_Allthreegaits}, $\tau_{zz}$ is smaller for the puller than for the neutral squirmer in the region immediately behind the body. 
One may thus expect the puller may swim faster than the neutral squirmer. However, large magnitude of $\tau_{zz}$ and $\tau_{rz}$ are observed in conjunction to the vortex ring, with  maximum values attained near the swimmer surface. Both components contribute to a polymeric (elastic) force applied on the swimmer as indicated by the grey arrows in  Fig.~\ref{fig:tzz_trz_al5_we0.5}. 
We observe an elongational flow   generated on top of the vortex ring,  and polymer chains stretched by such flow result in a strong force opposing the swimmer  motion, possibly explaining why the puller swimmer is not faster than the neutral squirmer despite displaying a significant reduction in polymer stretching behind its body.

To summarize, we observe elongational flows generated by all three swimmers 
around their bodies (neutral squirmer,  pusher, and puller).  The strength, orientation and position of the elongational flow is dependent on the swimming gaits. Such flows yield increasing elongational viscosities which serve as an additional elastic force, possibly positive or negative, to the  locomotion. For  the three gaits considered here, the force is  predominantly resistive, thus impeding locomotion. These computational results are in agreement with a recent experimental study of swimming \textit{Caenorhabditis elegans} in polymeric fluids where the authors suggest that  elongational viscosities can explain the measured decrease in swimming speeds~\cite{PRL_Shen_Visco}.

\section{Swimming power in viscoelastic fluids}

\begin{figure}
 \begin{center}
  \includegraphics[width=0.55 \textwidth]{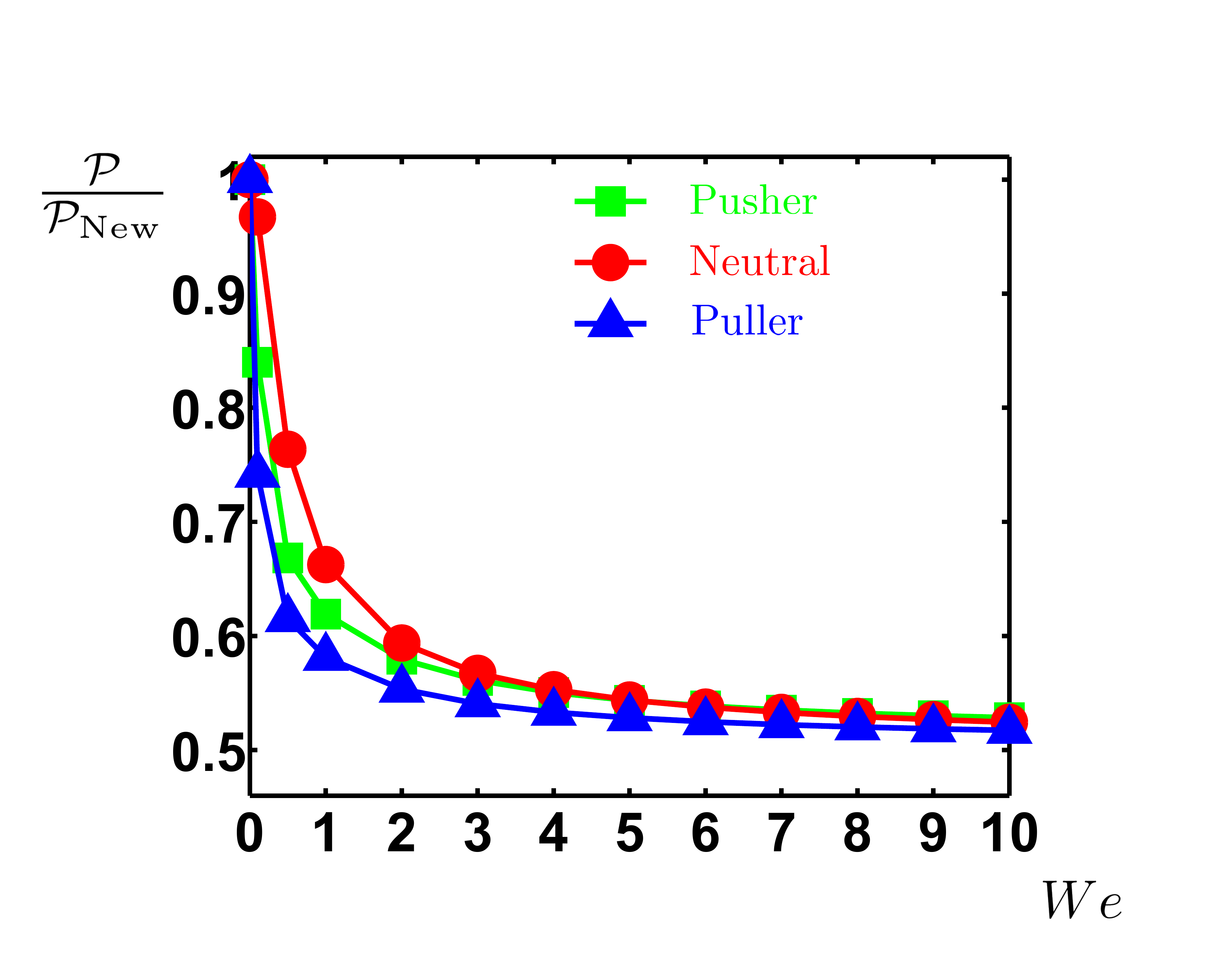}
 \end{center}
\caption{(Color online) Swimming power, $\mathcal{P}$, for pusher (squares, green online), puller  (triangles, blue
online) and neutral
(circles, red online) squirmers, nondimensionalized by the swimming power in the Newtonian
case,  $\mathcal{P}_{\rm New}$.The power consumption of swimming in the
Newtonian fluid are equal to  $\mathcal{P}_{\rm New}=8/3 \pi$ (neutral squirmer) and
 $\mathcal{P}_{\text{New}} = 36\pi$ (both pusher and puller).}
\label{fig:AR1_pow_we}
\end{figure}

The mechanical power, $\mathcal{P}$, expended by the spherical squirmer is
reported as a function of the Weissenberg number, $We$, for the three different
gaits under investigation  in Fig.~\ref{fig:AR1_pow_we}. When considering
dimensional quantities, we observe that both the puller and pusher require
significantly more power than the neutral squirmer due to the strong vortex
rings (see Fig.~\ref{fig:flow_pushpull}). Indeed, as reported in Ref.~\cite{Blake1971a},
$\mathcal{P}$ grows quadratically with the dipole magnitude $\alpha$. In order
to factor out the influence of the swimming gait on the  power expenditure,
thereby isolating  the effect of elasticity on the energy budget, we normalize
the power  with its corresponding value in the Newtonian fluid, termed  $\mathcal{P}_{\rm New}$.
In Fig.~\ref{fig:AR1_pow_we}, we observe a significant reduction in power in the
viscoelastic fluid for all gaits; $\mathcal{P}$ decreases first rapidly in the
range $We \in [0,2]$, and tends to  an asymptotic value when further increasing
the polymer relaxation time. The limiting value of $\mathcal{P}$ for $We\gg 1$
is close to half the value of the Newtonian case ($We=0$),
namely, ${\mathcal{P} |_{We\gg 1}}/{\mathcal{P}_{\text{New}}} \approx 0.5$.
The result is in agreement with the asymptotic analysis of Ref.~\cite{lauga07}  which predicts 
 ${\mathcal{P} | _{We\gg 1}}/{\mathcal{P}_{\text{New}}} =
\beta$ ($\beta=0.5$ for the computations presented here).

As we will show below, the component of the power consumption associated with the Newtonian
solvent is almost constant  with variations in $We$, while the contribution of
the polymeric stresses decreases significantly (up to one order of magnitude
decrease at large $We$). Since, for $\beta=0.5$, the Newtonian and polymeric
contribution are the same for $We=0$, the power at large Weissenberg numbers is
about half that in a Newtonian fluid.

To probe  the power saving for microswimmers in viscoelastic fluids, we decompose the power into three parts. First,
we use the divergence theorem to transform the total power, $\mathcal{P}$, which is the integral of the stress at the
surface times the swimming speed,  into a volume integral as done in \cite{PRLswim}
\begin{equation}
 \mathcal{P}=-\int_S \mathbf u \cdot {\boldsymbol{\tau}} \cdot \mathbf n \, dS = \int_V \nabla \cdot (\mathbf u
\cdot \boldsymbol{\tau}) \, dV,
\end{equation}
where $\mathbf{n}$ is the unit normal outward the swimmer surface $S$, $\mathbf{u}$ is the velocity in the laboratory
frame and $\boldsymbol{\sigma}$ is the stress tensor introduced above.
Rearranging the integrand, one can write
\begin{equation}
 \nabla \cdot (\mathbf{u} \cdot \boldsymbol{\boldsymbol{\tau}}) = {\beta\mu
\boldsymbol{\omega}^{2}}+2\beta\mu
(\nabla{\mathbf{u}}:\nabla{\mathbf{u}})+\mathbf{E}: \boldsymbol{\tau^p} .
\end{equation}
Consequently, $\mathcal{P}$ is found to be the sum of three contributions,
$\mathcal{P}=\mathcal{P}_{\Omega}+\mathcal{P}_{DV}+\mathcal{P}_{P}$: a component related to the flow vorticity, 
$\mathcal{P}_{\Omega}=\int_V{\beta\mu \boldsymbol{\omega}^{2}}dV$, a component related to the velocity gradient,
$\mathcal{P}_{DV}=\int_V 2\beta\mu (\nabla{\mathbf{u}}:\nabla{\mathbf{u}})dV$, and a polymeric component,
$\mathcal{P}_{P}=\int_V{\mathbf{E} :  \boldsymbol{\tau}^{p}}dV$, where $\mathbf{E}$ is the rate-of-strain
tensor.
The component related to the velocity derivative, $\mathcal{P}_{DV}$, is only dependent on the tangential velocity
distribution we impose on the surface and thus is only a function of $\alpha$. This can be seen by re-writing
\begin{equation}\label{eq:vederi}
 \mathcal{P}_{DV}= 2\beta\mu \int_V  (\nabla{\mathbf{u}}):(\nabla{\mathbf{u}})  \, dV =-2
\beta \mu \int_{S}
\mathbf{n} \cdot \ ({\mathbf{u}} \cdot \nabla {\mathbf{u}} ) \, dS.
\end{equation}

\begin{figure}[h]
   \centering
   \includegraphics[width=0.48 \textwidth]{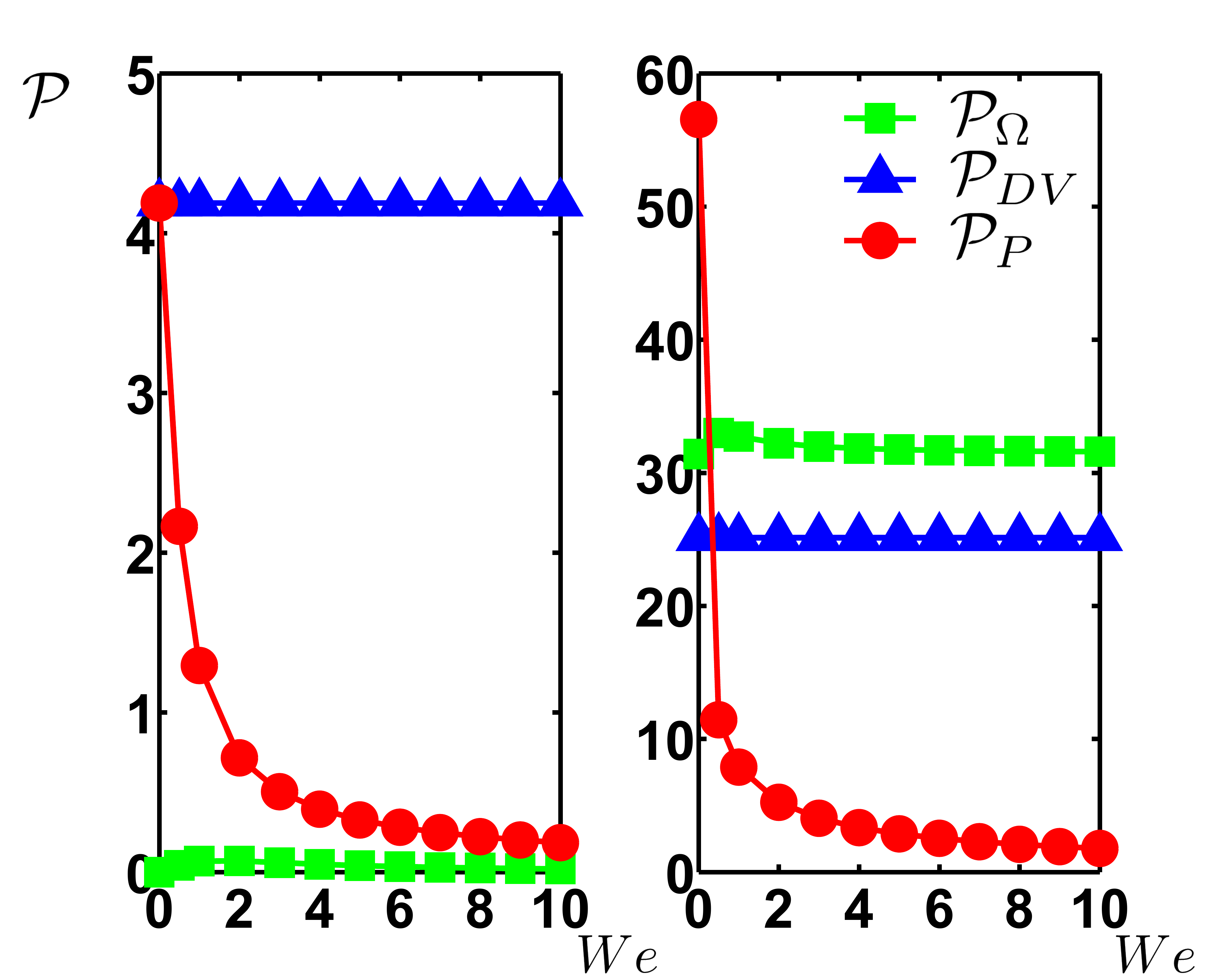}
  \caption{(Color online) Contributions to the total  power expended by swimming, $\mathcal{P}$, versus Weissenberg
number $We$, $\beta=0.5$, for neutral squirmer
(left) and puller (right). The three contributions are defined as: $\mathcal{P}_{\Omega}=\int_V{\beta\mu
\boldsymbol{\omega}^{2}}dV$,
$\mathcal{P}_{DV}=\int_V 2\beta\mu (\nabla{\mathbf{u}}:\nabla{\mathbf{u}})dV$ and $\mathcal{P}_{P}=\int_V{\mathbf{E} :
\boldsymbol{\tau}^{p}}dV$.}
   \label{fig:powdecomp}
\end{figure}

The three different contributions to the total consumed power are displayed in
Fig.~\ref{fig:powdecomp} as a function of the Weissenberg number, $We$, for two
swimming gaits, neutral swimmer (left panel) and puller (right panel). Energetic
results for pushers are qualitatively similar to those for pullers. 
First note that for $We=0$ and our choice $\beta=0.5$, the power component due
to the polymeric stress is $\mathcal{P}_{P}=\mathcal{P}_{\Omega} +
\mathcal{P}_{DV}$. As seen from the definition of the polymeric model,
$\boldsymbol{\tau^p}$ reduces to Newtonian stress $\ (1-\beta\ )(\nabla
\mathbf{u} + \nabla \mathbf{u}^{T})$ when $We=0$ in  Eq.~\eqref{eq:poly}.

For a fixed mode of surface motion, $\mathcal{P}_{DV}$,  is strictly constant  with $We$ since it depends only on the
values of the velocity at the surface (see Eq.\ref{eq:vederi}).  For both gaits, we observe computationally that the
vorticity contribution, $\mathcal{P}_{\Omega}$,  shows little variations with $We$.  The major contribution to the
reduction in swimming power stems thus from the significant reduction of the polymeric part, $\mathcal{P}_{P}$, with
increasing values of the Weissenberg number.

The contribution of polymeric stresses to the consumed power is the double dot product of two tensors, the
rate-of-strain
tensor, $\mathbf{E}=\frac{1}{2}(\nabla{\mathbf{u}}+(\nabla{\mathbf{u}})^{{T}})$, and the polymeric stress tensor,
$\boldsymbol{\tau}^{p}$. Polymer chains get stretched by the rate-of-strain. In fluids with low relaxation times,
polymeric stretching occurs immediately, resulting in high spatial correlation between the polymeric stress tensor and
the rate-of-strain tensor. Thus, the product of the two tensors takes large values. In contrast, for fluids with large
relaxation times, polymer chains get fully stretched at a position far away from where they get excited, and the spatial
correlation of the two tensors decreases. Polymeric deformation and rate-of-strain become therefore separated in space
as $We$ increases, which gives a systematically smaller power consumption, as observed computationally.

\begin{figure}[h]
   \centering
   \includegraphics[width=0.4 \textwidth]{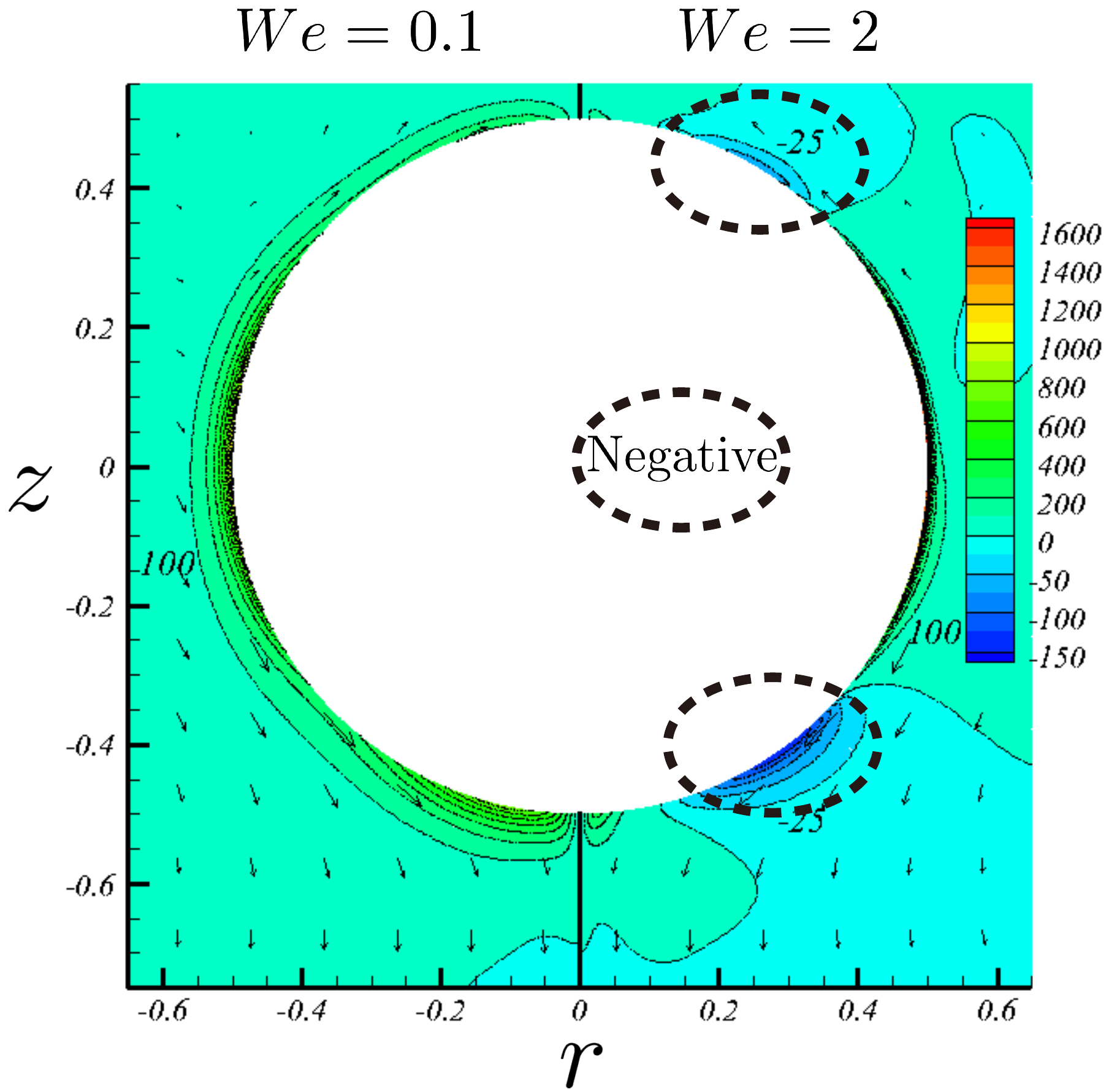}
   \caption{(Color online)  Polymeric power density, $D_{P}$, for pushers in
fluids with $We=0.1$ (left) and $We=2$ (right). Dotted elliptical circles mark
areas with negative values of $D_{P}$. The values of $D_{P}$ along the
swimmer surface are displayed in Fig.~\ref{fig:powpoly_surf}.}
   \label{fig:powpoly}
\end{figure}

\begin{figure}[b]
   \centering
   \includegraphics[width=0.5 \textwidth]{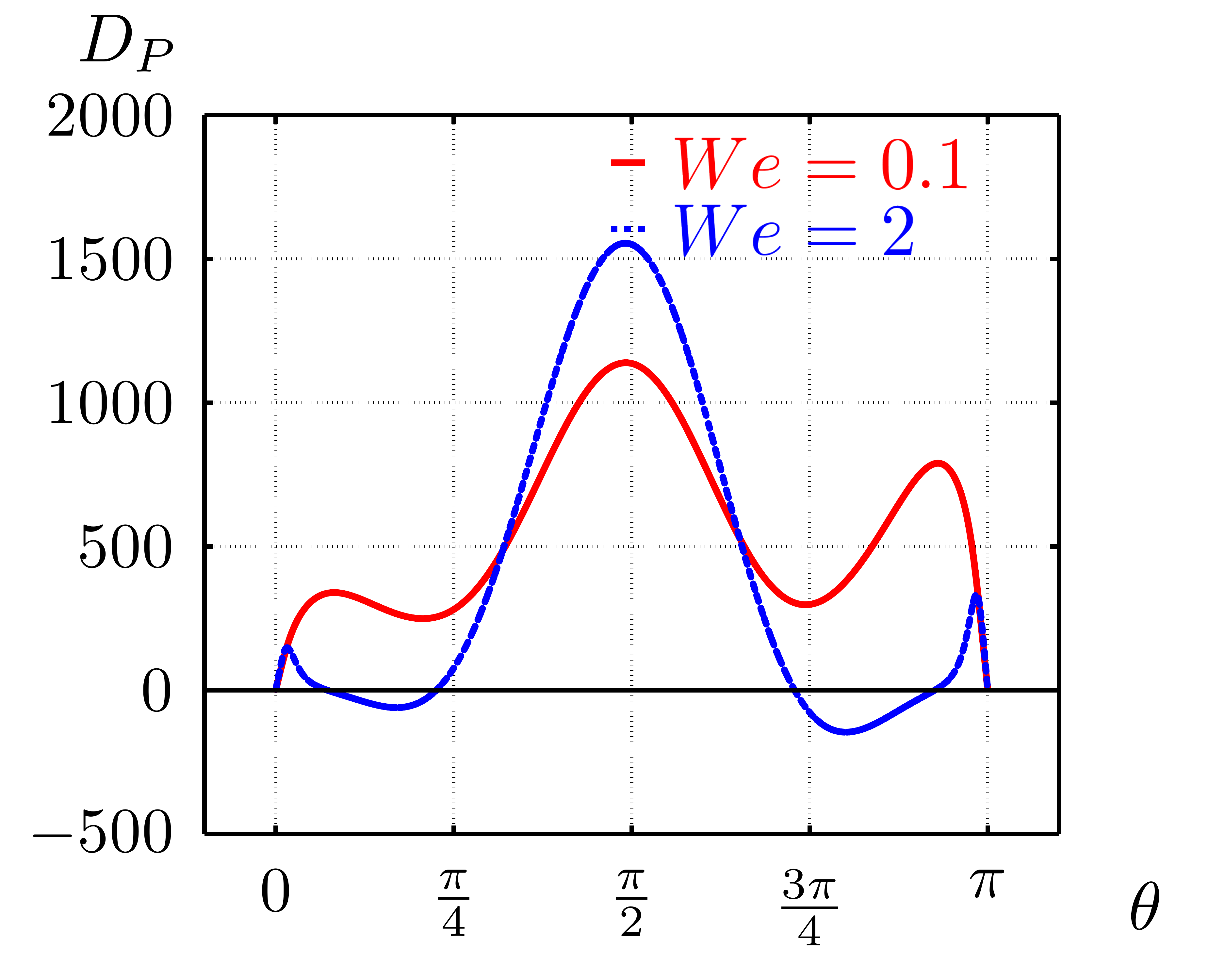}
   \caption{(Color online) Spatial distribution of polymeric power density,  $D_{P}$, along the surface of
pushers in viscoelastic fluids with $We = 0.1$ (solid line, red online) and $2$
(dashed line, blue online); $\theta$ is defined in Sec.~\ref{mathmodel}, as the angle between
the swimming direction and the position vector of each point.}
   \label{fig:powpoly_surf}
\end{figure}

In Fig.~\ref{fig:powpoly}, we plot the power density, $D_{P}=2 \pi |r|
\mathbf{E}:\boldsymbol{\tau^p}$, for the spherical pusher for two different
values of $We$.  When $We=0.1$ (left half of the figure), large values of
$D_{P}$ are observed in the region very close to the swimmer, especially in
the equatorial region. This is the region where most of the energy is expended to
elongate the polymers. A region of positive $D_{P}$ also exists near the
equatorial $z=0$ plane for the case with $We=2$ shown on the right, although of
reduced size. In addition, negative density can be seen on the planes $z \approx
\pm 0.4$ (indicated by elliptical circles in the figure), which can  also be clearly
identified in Fig.~\ref{fig:powpoly_surf} where we  display the value of
$D_{P}$ evaluated along the surface of the pusher. These areas of negative
power density produce further decrease of the total expended power by providing
energy to the system. This energy analysis allows us to  understand the
reduction in swimming power observed in viscoelastic fluids for any type of
surface deformation considered.

\begin{figure}[t]
 \centering
   \includegraphics[width=0.4 \textwidth]{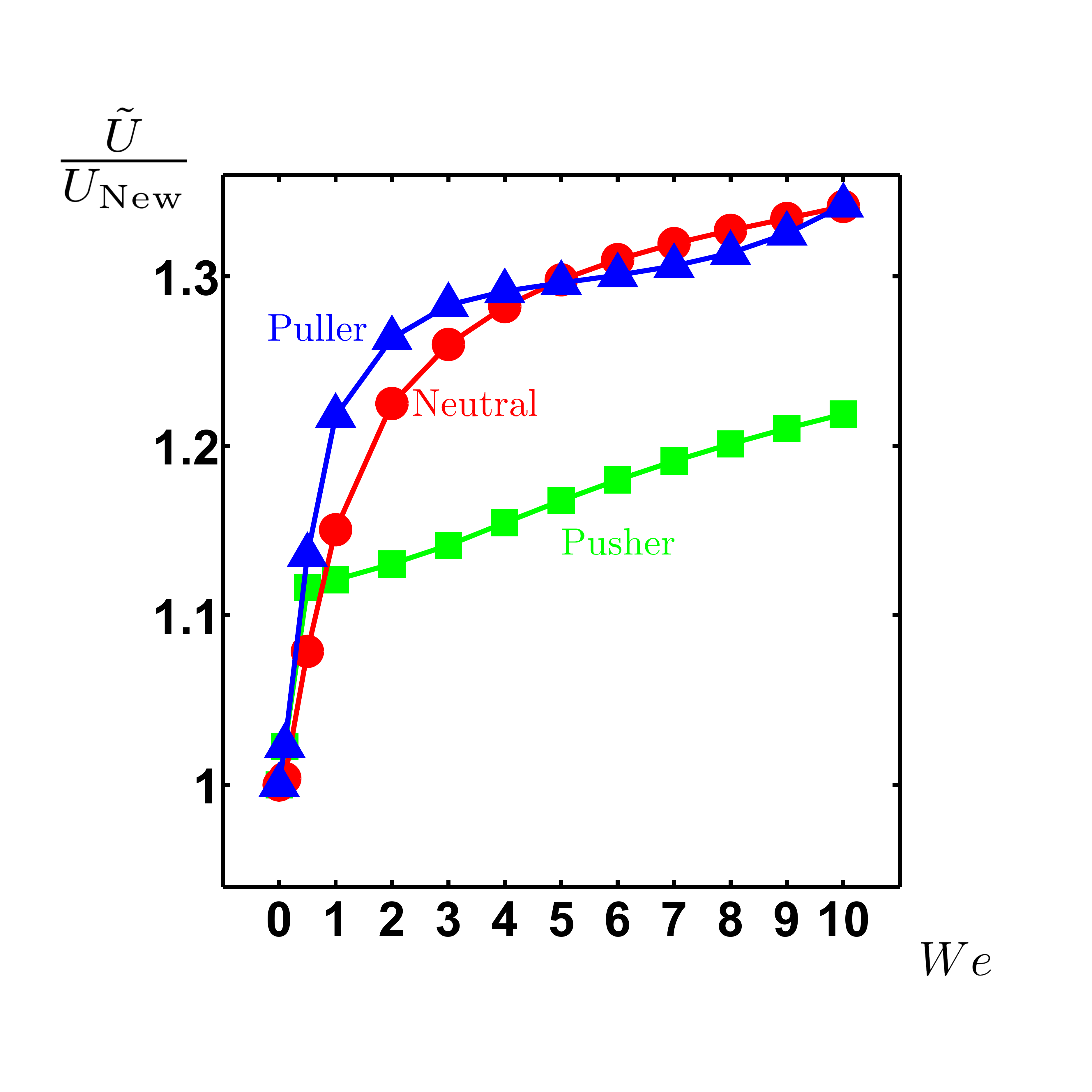}
  \caption{Swimming speed $\tilde U$, scaled to ensure locomotion at constant power, versus Weissenberg  number, $We$,
for $\beta=0.5$. Nondimensional form $\frac{\tilde U}{U_\text{New}}$ is plotted for the three swimming
gaits: neutral (circles, red online), puller (triangles, blue online) and pusher (squares, green
online)}.
   \label{fig:CONpow_spd}
\end{figure}

We finally consider the situation of locomotion at constant power. 
Some bacteria have been observed to swim with kinematics consistent 
with constant-power conditions~\cite{swimConstantpower}, and it is thus  interesting to consider the swimming speed
which would be attained
by  squirmers if they were under the constraints of expending the  same power as
in a Newtonian fluid (the  results in Fig.~\ref{fig:AR1_spd}
were obtained at constant intensity of the velocity at the boundary). The
swimming speed obtained rescaling the previous data to obtain constant-power
locomotion is displayed in Fig.~\ref{fig:CONpow_spd}.  
To calculate the velocity at constant power, we need to increase the value of the surface velocity $B_1$ to account for the power-saving in polymeric fluid discussed above. If we use $B_{1}^{\prime}$ to denote
the new value of $B_{1}$, constant power requirement is written as
$\left(B_{1}^{\prime}\right)^{2}\mathcal{P}|_{We\left(B_{1}^{\prime}\right)} =
B_{1}^{2}\mathcal{P}_{\text{New}}$, where $We\left(B_{1}^{\prime}\right) = \frac{\lambda
B_{1}^{\prime}}{D} = \frac{B_{1}^{\prime}}{B_{1}} We$; we   use a simple interpolation from our computational  results in order to obtain 
$\mathcal{P}|_{We\left(B_{1}^{\prime}\right)}$. Thus, the swimming speed with constant
power is computed as $\tilde{U} =
\frac{B_{1}^{\prime}}{B_{1}}U|_{We\left(B_{1}^{\prime}\right)}$. As can be seen in Fig.~\ref{fig:CONpow_spd},  $\tilde{U}$ is found to be systematically larger than that in the Newtonian
conditions for all the cases considered, and the swimming speed systematically increases with the Weissenberg  number. The puller and the neutral squirmer are about 20\% faster than the pusher.

\section{Swimming Efficiency  in viscoelastic fluids}

\begin{figure}[t]
   \centering
   \includegraphics[width=0.48 \textwidth]{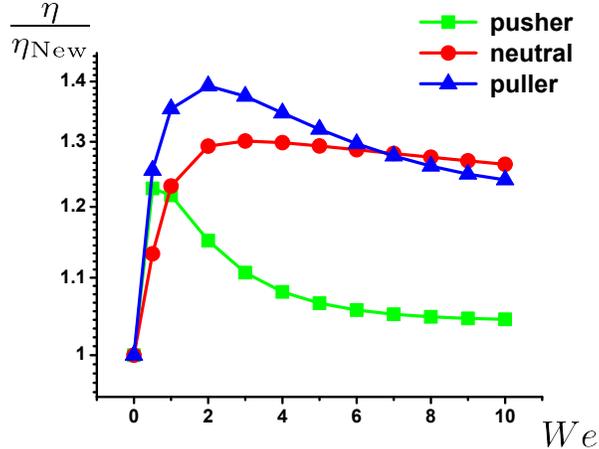}
  \caption{(Color online) {Swimming efficiency, ${\eta}$, normalized by the
corresponding value for  Newtonian swimming,  ${\eta_{\text{New}}}$,
versus Weissenberg number, $We$. The efficiencies of swimming in the
Newtonian fluid are equal to  $\eta_{\text{New}}=0.5$ (neutral squirmer) and 
 $\eta_{\text{New}}=0.037$ (both pusher and puller).}}
   \label{fig:eff}
\end{figure}

The  hydrodynamic efficiency of squirmers in viscoelastic fluid is displayed in
Fig.~\ref{fig:eff}. The  efficiency is defined as the ratio between the rate  work
needed to pull a sphere of same size in the same fluid  at the swimming speed $U$ and
the swimming power $\mathcal{P}$ of a self-propelled swimmer \cite{lp09,effoptiMichelin} 
\begin{equation}{\label{effdefine}}
 \eta = \frac{F U}{\mathcal{P}},
\end{equation}
where $F$ is the force required to drag the spherical squirmer body at the speed $U$.  Similarly to what we did for the power consumption, the efficiency is normalized in Fig.~\ref{fig:eff} by its Newtonian value. We observe that all swimmers have a higher swimming efficiency  than in a Newtonian fluid. 
For all three gaits, as $We$ increases from $0$, the scaled efficiency firstly increases rapidly with $We$, then decreases asymptotically. We can thus identify an optimal $We$ number corresponding to the maximum efficiency; this is around $2$ for the neutral and puller swimmer and around $0.5$ for the pusher. 

\section{Velocity decay  in viscoelastic fluids}

\begin{figure}[t]
   \centering
   \includegraphics[width=0.45 \textwidth]{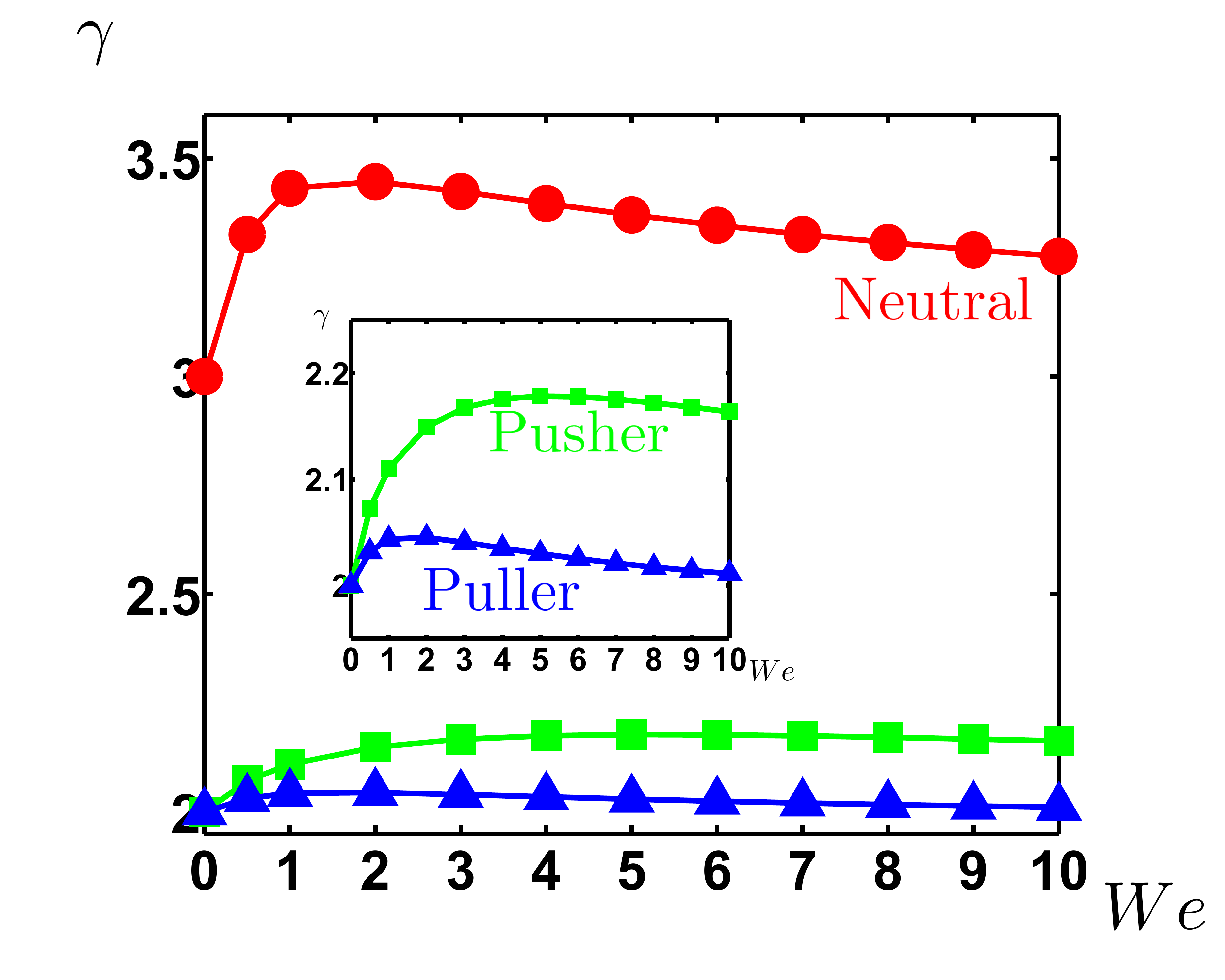}
  \caption{(Color online) Power law exponent, $\gamma$, for spatial decay of the axial  velocity along
the axis of the domain ($r=0$) ($|u| \sim r^{-\gamma}$, see text), as a
function of the Weissenberg  number, $We$.  The inset plot
highlights the difference between pusher and puller.}
   \label{fig:decay}
\end{figure}

Because of its implications to  collective cell behavior
\cite{lp09,PRL_Shen_Visco}, it is of interest to investigate the effect of the fluid elasticity on the
spatial decay of the flow perturbation induced by the swimming motion. 
We compute the axial velocity along the symmetry
axis of the domain for the pusher and puller (similar analysis was carried out
for the neutral swimmer in {our previous
paper}~\cite{lailai-pre}). The velocity decays with a power-law behavior,
$|u|\sim 1/r^{\gamma}$, with $\gamma=2$ in the Newtonian case for both pushers
and pullers, as sufficiently far away from the cell the higher order terms
($1/r^{3}$ and $1/r^{4}$) give negligible contribution to the flow field ($\gamma=3$ in the neutral
potential-flow case). In the
non-Newtonian case, we estimate the value of $\gamma$ by fitting a power law to
the numerical results in a measurement region extending from about $r \approx
5D$ to the end of the computational domain. The estimated values of $\gamma$ are
displayed in Fig.~\ref{fig:decay} as a function of the Weissenberg number,
$We$. We see clearly that the velocity always decays faster in viscoelastic fluids than in the
Newtonian fluid. For both pusher and puller, $\gamma$ is not monotonically
increasing when with $We$, but instead reaches a maximum value at intermediate
Weissenberg numbers. The maximum value of $\gamma$ is seen to take place as $We$
is around $5$ for the pusher and around $1$ for the puller, which approximately
corresponds with the values of the Weissenberg numbers at which the swimming
speed, $U$, is minimum for both gaits (although slightly above). We thus observe
a negative correlation between the swimming speed
and the velocity decay rate. We further note that the velocity decay rate  is
larger for the pusher than for the puller, which is  consistent with our
observation of the negative correlation as the swimming speed of a pusher is
smaller than that of a puller.

\section{Stresslet  in viscoelastic fluids}

In this final section we address how the viscoelastic modification of the squirming motion would affect the rheology of
an active suspension of squirmers. When the size of the microorganisms is much smaller than the scale of the flow field
under consideration, it is convenient to use a continuum model of the active suspension. The first order correction to
the bulk viscosity in terms of the volume fraction of swimmers is then given by the stresslet associated with an
individual swimmer. Batchelor~\cite{Batchelor_Stress70} derived a relation between the bulk stress and the conditions at
the surfaces of individual particles embedded in a Newtonian solvent (his original formulation is given in dimensional
form and we present the nondimensional version here using the scaling we introduced in Sec.~\ref{mathmodel}). The 
bulk stress $\boldsymbol{\hat{\sigma}}$ is given by
\begin{equation}
 \boldsymbol{\hat{\sigma}} = \bold{I.T.} + 2\bar{\bold{E}} +
\boldsymbol{\sigma^{\prime}},
\end{equation}
where $\bold{I.T.}$ is
an isotropic tensor, i.e.\ $\bold{I.T.} =
\hat{\sigma}_{kk}\bold{I}$,
 {${\bar{\bold E}}$ is the volume-averaged rate-of-strain tensor ($\bar{E}_{ij} =
\frac{1}{2V}\int_{V}\left(\frac{\partial u_{i}}{\partial x_{j}}+\frac{\partial u_{j}}{\partial x_{i}} \right)dV$,
where $V$ is the volume occupied by the fluid and the particles) and
$\boldsymbol{\sigma^{\prime}}$ represents the disturbance stress due} to the
presence of  squirmers; see also Ref~\cite{Ishi_2headhelix} for the application to active suspensions in Newtonian
fluids.
The particle bulk stress $\boldsymbol{\sigma^{\prime}}$
induced by force- and torque-free particles (such as swimmers)
in the volume $V$ can be expressed as
\begin{eqnarray} \label{eq:stress}
 \boldsymbol{\sigma^{\prime}} &=& \frac{1}{V}\sum{\mathbf{S}},\\ 
 \mathbf{S} &=& \int_{A_{S}} \left[\frac{1}{2}\left\{\left(
\boldsymbol{\sigma}\cdot \mathbf{n} \right)\mathbf{x} + \mathbf{x}\left(
\boldsymbol{\sigma}\cdot \mathbf{n} \right) \right\} -
\frac{1}{3}\mathbf{x} \cdot \boldsymbol{\sigma} \cdot \mathbf{n}\mathbf{I} -
\left(\mathbf{u}\mathbf{n} + \mathbf{n}\mathbf{u} \right)\right]dA.
\label{def:stresslet}
\end{eqnarray}
The stress is thus given by the volume average of the so-called stresslet $\mathbf{S}$, where the sum is taken over all the swimmers. The stresslet $\mathbf{S}$ is defined Eq.~\eqref{def:stresslet} where $\boldsymbol{\sigma}$ is the stress tensor, $A_{S}$ is the surface of one of the squirmers, $\mathbf{n}$ is the unit outward normal
vector, $\mathbf{x}$ is the position vector and $\mathbf{I}$ is the unit tensor.

Ishikawa and coworkers \cite{swimInter-Pedley} exploited  the formulation above to study the stresslet $\mathbf{S}$ of
a single squirmer and a pair of squirmers in the Newtonian fluid both analytically and
numerically. They showed that for single squirmers
\begin{equation}\label{eq:stressletAna}
 \mathbf{S} = \frac{1}{3}\pi\left(3\mathbf{e}\mathbf{e} - \mathbf{I} \right)B_{2},
\end{equation}
where 
$\mathbf{e}$ is the swimming direction and $B_{2}$ the second squirming mode defined in Sec.~\ref{mathmodel}. The assumption of axisymmetric bodies makes the off-diagonal part of
$\mathbf{S}$ zero. In a Newtonian fluid, the first mode $B_1$ determines the swimming speed while  $B_{2}$  the stresslet magnitude; the stresslet of a neutral squirmer ($B_{2}=0$) is therefore zero.

In order to derive a similar formalism for a squirmer in a non-Newtonian flow, we consider the additional polymeric contribution, 
$\boldsymbol{\tau}^{p}$, to the Newtonian stress, $\boldsymbol{\sigma} = -p\mathbf{I} +
\beta\left(\nabla \mathbf{u} + \nabla \mathbf{u} ^{T}\right) + \boldsymbol{\tau}^{p}$. The same derivation as in Refs.~\cite{Batchelor_Stress70,Batchelor_Stress72} can be performed for non-Newtonian fluids, see for example Ref.~\cite{stressletViscoelastic}. In our case, 
we seek an expression for $\boldsymbol{\sigma^{\prime}}$
\begin{equation}\label{stresslet1}
 \boldsymbol{\hat{\sigma}} = \bold{I.T.} + 2\beta\bar{\bold{E}} + \boldsymbol{\bar \tau_p} +
\boldsymbol{\sigma^{\prime}},
\end{equation}
where $\boldsymbol{\bar \tau_p}$ { is the volume average of the polymeric stress ($\boldsymbol{\bar \tau_p} =
\frac{1}{V}\int_{V}\boldsymbol{\tau_p}dV$).
The total bulk stress can be written as \cite{Batchelor_Stress70}
\begin{equation}\label{stresslet2}
\boldsymbol{\hat{\sigma}} = \frac{1}{V}\int_{V-V_{S}}\{-p\mathbf{I} + \beta\left(\nabla \mathbf{u} + \nabla
\mathbf{u}^{T}\right) + \boldsymbol{\tau_p}\}dV + \frac{1}{V}\int_{V_{S}}
\boldsymbol{\sigma}dV,
\end{equation}
where $V_S$ is the volume occupied by the squirmer.}
Combining Eq.~\eqref{stresslet1} and Eq.~\eqref{stresslet2}, we have
\begin{equation}\label{str3}
 \boldsymbol{\sigma^{\prime}} = -\frac{1}{V}\int_{V_{S}}\beta\left(\nabla \mathbf{u} + \nabla
\mathbf{u}^{T}\right)dV - \frac{1}{V}\int_{V_{S}}\boldsymbol{\tau_{p}}dV + \frac{1}{V}\int_{V_{S}}\boldsymbol{\sigma}dV.
\end{equation}
{The volume integral of the Newtonian contribution to the total stress inside the squirmer is 
re-written as a surface integral by applying the divergence theorem: the first term on the right hand side of Eq.~\eqref{str3}
becomes  $-\frac{1}{V}\int_{A_{S}}\beta\left(\mathbf{u}\mathbf{n} + \mathbf{n}\mathbf{u}\right)dA$, whereas  the total stress inside the particles contributes to the stresslet
\begin{equation}\label{aversig}
\int_{V_{S}}\boldsymbol{\sigma}dV = \int_{A_{S}}\left(\boldsymbol{\sigma} \cdot \mathbf{n}\right)\mathbf{x}dA -
\int_{V_{S}}\left(\nabla \cdot \boldsymbol{\sigma}\right) \mathbf{x}dV,
\end{equation}
where we have assumed $\nabla \cdot \boldsymbol{\sigma} = 0$ inside the solid body.} For a swimming squirmer which is
force-free and torque-free, $\left(\boldsymbol{\sigma} \cdot \mathbf{n}\right)\mathbf{x} = \frac{1}{2}\left\{\left(
\boldsymbol{\sigma}\cdot \mathbf{n} \right)\mathbf{x} + \mathbf{x}\left(
\boldsymbol{\sigma}\cdot \mathbf{n} \right) \right\} - \frac{1}{3}\mathbf{x} \cdot \boldsymbol{\sigma} \cdot
\mathbf{n}\mathbf{I}$ as in Ref.~\cite{swimInter-Pedley}.
Finally we obtain the new definition of the stresslet
\begin{equation} \label{eq:stress_non}
 \mathbf{S} = \int_{A_{S}} \left[\frac{1}{2}\left\{\left(
\boldsymbol{\sigma}\cdot \mathbf{n} \right)\mathbf{x} + \mathbf{x}\left(
\boldsymbol{\sigma}\cdot \mathbf{n} \right) \right\} -
\frac{1}{3}\mathbf{x} \cdot \boldsymbol{\sigma} \cdot \mathbf{n}\mathbf{I} -\beta
\left(\mathbf{u}\mathbf{n} + \mathbf{n}\mathbf{u} \right)\right]dA -\int_{V_S} \boldsymbol{\tau}^{p} \, dV.
\end{equation}

\begin{figure}[b]
   \centering
   \includegraphics[width=0.55 \textwidth]{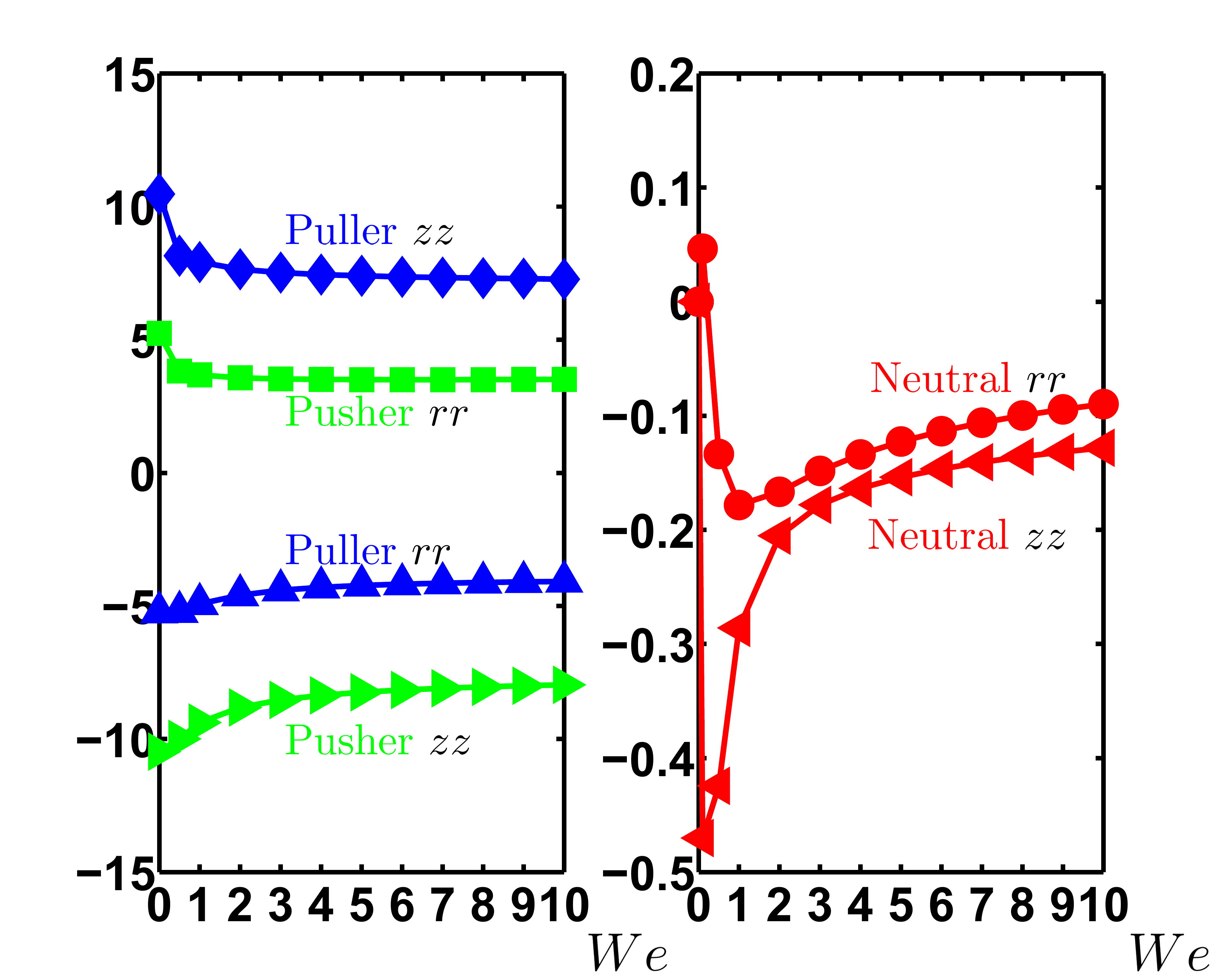}
  \caption{(Color online) Stresslets induced by swimmers in viscoelastic fluids:   $rr$ and $zz$ components of the tensor $\mathbf{S}$, Eq.~\eqref{eq:stress_non},  for different swimming gaits.
Left: pusher ($\alpha=-5$) and puller ($\alpha=5$). Right: neutral
squirmer ($\alpha=0$).}
   \label{fig:stresslet}
\end{figure}

The stresslets $\mathbf{S}$ computed from Eq.~\eqref{eq:stress_non} for the pusher ($\alpha=-5$), puller ($\alpha=5$) and neutral squirmer ($\alpha=0$) are displayed in Fig.~\ref{fig:stresslet}. Because of  axisymmetry, there are only two non-zero components of $\mathbf{S}$, namely, the $rr$ component in the direction normal to the swimming direction and the $zz$ component in the swimming direction.

First, we see that for the pusher and puller the magnitude of both the $rr$ and $zz$ components decreases monotonically with increasing $We$, and reaches an asymptotic value (while keeping the same sign). This corresponds to a maximum decrease of approximately $20\%$ with respect to the Newtonian value at $We=10$. This implies that a dilute suspension of swimming cells will have a weaker rheological effect in a viscoelastic fluid. 

The decrease in the stresslet can tentatively be explained by the fact that the
polymeric stress always opposes the flow which generates it. Let us consider the
$zz$ component of a pusher as an example; in that case, the tangential velocity
on the surface  pushes the fluid away from the cell along  the swimming
direction as shown in Fig.~\ref{fig:flow_pushpull}, resulting in an extensional
flow in front and behind the cell body with a consequent negative $zz$ component
of the stresslet. Simultaneously, the extensional flow produces strong positive
polymeric stresses
(see Fig.~\ref{fig:tzz_al-5_al5_we01_2}) counteracting the effect created by the
wall actuation, and hence decreasing the net stress; the  impact on the $rr$ component can be explained in a similar
fashion. Note that for both the pusher and puller, the magnitude of the
$zz$ component is approximatively twice that of the $rr$ component, as can be seen from  Eq.~\eqref{eq:stressletAna}.

Interestingly, we also observe in Fig.~\ref{fig:stresslet} that for the neutral squirmer, which has zero stresslet in the Newtonian limit,  viscoelastic stresses induce a non-zero stresslet. The value of the $rr$ component is lower than that of the $zz$ component and of same sign. 
Considering two squirmers swimming in the same direction side by side, at leading order in their separation distance they  have no influence on each other in a Newtonian fluid, but will repel each other due to the negative value of the $rr$ component of the stresslet. Similarly,  two neutral squirmers swimming in line along the same swimming direction will   tend to increase their relative distance owing to the  negative value of the $zz$ component of the stresslet. The peak value of the $rr$ and $zz$ components occur at $We \approx 1 $, at which the minimum swimming speed of the neutral squirmer appears, as well as the fastest velocity decay. Interestingly, this viscoelastic repulsion between  swimming squirmers has the sign  opposite to the viscoelastic attraction observed between sedimenting spheres in polymeric fluids.

\section{Conclusion}

In the current paper, we numerically compute the swimming speed, power, and hydrodynamic efficiencies of model ciliated cells swimming in a viscoelastic Giesekus fluid. We use the squirmer model where the self-propulsion is driven by tangential surface deformation and consider three types of boundary conditions to mimic swimming gaits typical of cell locomotion (potential squirmer, pusher, and puller) all having the same swimming speed in a Newtonian fluid. The characteristics of the flow and of the polymeric stress are examined to help understand our computational results. 

For constant magnitude of the surface deformation velocity, we show that the swimming speed decreases in the viscoelastic fluid compared its the Newtonian value for all cases considered, with a minimum obtained at intermediate values of the Weissenberg number. Neutral swimmer and puller have the highest velocity, whereas the pusher is about 15\% slower.  For the potential squirmer and the  pusher, the reduced self-propulsion is explained by the accumulation of highly stretched polymers behind the body.  For low values of the polymer relaxation times, $We \lesssim 1$, the stretching of the polymers increases hence the initial decrease in the swimming speed. For the largest values of the Weissenberg number, however, the region of stretched polymers behind the swimmer becomes thinner and the induced elastic resistance decreases, thus explaining the slow recovery of the swimming speed.  In the case of pullers, a  different velocity field is created and stretched polymers are advected away from the cell on the sides. This induces a force with a component both in the radial and axial direction, the latter explaining the reduced swimming speed. Similar kinematic  analysis  could be used to  provide detailed guidelines for the design of efficient swimmers in viscoelastic fluids -  the basic design principle consisting in reducing the accumulation of stretched polymers at the cell surface in the direction of motion.

Along with the decrease in swimming speed, we  observe  a significant reduction of swimming power in all cases. To understand this observation, we analytically decompose the expression for the power expended by cells swimming in viscoelastic fluids into three contributions (integrals over the flow domain): one associated to the vorticity induced in the flow, one only set by the surface deformation, and one related to the polymeric stresses.  In this way, we are able to clearly identify the reduction in consumed power with a reduction in the polymeric contribution and interpret the  observed  reduction in power  as due to an increasing spatial de-correlation between the flow shear and the induced polymer deformation. In addition, we observe regions of negative power density  indicating that the polymers in the fluid have the capacity to first transform the mechanical energy into potential  energy (polymer deformation), and then feed it back to the fluid thus reducing the consumed power. We also consider the hydrodynamic efficiencies and show that self-propulsion is always more efficient 
in a viscoelastic fluid, with a maximum relative efficiency obtained for intermediate values of the Weissenberg number.

We further investigate the influence of the fluid elasticity on the decay rate of the perturbation velocity induced by the squirmer. We find that the decay rate is faster in viscoelastic fluid and is larger for the pusher than for the puller.  Finally, to address the influence of viscoelasticity on the rheology of an active suspension of swimmers, we extract from the
computational results the stresslet associated with an isolated swimmer. Notably, viscoelasticity  induces a non-zero stresslet for the potential squirmer (rigorously equal to zero in the Newtonian case).  As a difference, the magnitude of the stresslet for both  pushers and pullers decrease by about $20\%$. These results presents a first quantitative step toward an understanding of the dynamics and rheology of active suspensions in viscoelastic fluids.

\section*{Acknowledgments}
Funding  by VR (the Swedish Research Council) to L.B., by the Linn\'e Flow Centre at KTH, and the US National Science Foundation (grant  CBET-0746285 to E.L.) is gratefully acknowledged.
Computer time provided by SNIC (Swedish National Infrastructure for Computing) is also acknowledged.

\bibliographystyle{unsrt}
\bibliography{refs}

\end{document}